\documentclass[prd,nofootinbib,12pt,amsart]{revtex4}
\usepackage[utf8]{inputenc}
\usepackage{lineno,hyperref}
\usepackage{amsfonts,amssymb,amsmath}
\usepackage{epsfig}
\usepackage{mathrsfs}
\usepackage{amssymb}
\usepackage{graphicx}
\usepackage{amsmath}
\usepackage{xcolor}
\usepackage{color}
\usepackage{appendix}
\usepackage{amsmath}

\usepackage{xcolor} 
\begin{document}

\title{A Bogoliubov-ratio framework for quantum-information diagnostics of time-dependent two-mode Boson System}
\author{Shi-Cheng Liu, Lei-Hua Liu}
\email[Corresponding author: ]{liuleihua8899@hotmail.com }
\author{Bichu Li}
\email{libichu@mail.ustc.edu.cn}
 \affiliation{Department of Physics, College of Physics,Mechanical and Electrical Engineering,Jishou University, Jishou 416000, China }
 \author{Yuebing Zhou}
 \email[Corresponding author: ]{zyb@hhtc.edu.cn}
 \affiliation{Department of Physics, Huaihua University, Huaihua, Hunan 418008, China}
\author{Hai-Qing Zhang}
\email{hqzhang@buaa.edu.cn}
\affiliation{Center for Gravitational Physics, Department of Space Science, Beihang University, Beijing, 100191, China }
\affiliation{Peng Huanwu Collaborative Center for Research and Education, Beihang University, Beijing
	100191, China }

\begin{abstract}
  We formulate a compact dynamical representation of quantum-information diagnostics for time-dependent two-mode	bosonic systems in terms of the Bogoliubov ratio $\lambda_k(\eta)=\beta_k(\eta)/\alpha_k(\eta)$.	For vacuum-evolved pure Gaussian pair states generated by a	Hermitian quadratic Hamiltonian, $\lambda_k$ obeys a closed complex Riccati equation. After one partner mode is traced out, the reduced-state spectrum is determined by $q_k=|\lambda_k|^2$, from which the purity, linear entropy,	R\'enyi-2 entropy, and von Neumann entropy follow directly. These Gaussian results are mathematically equivalent to those obtained from the standard density-matrix or covariance-matrix formalism; the advantage of the ratio representation is a compact separation between model-dependent dynamics and universal state diagnostics. Writing $\lambda_k=\sqrt{q_k}e^{i\theta_k}$ further exposes how pair production drives radial growth, whereas frequency rotation and detuning act through the phase and can suppress coherent squeezing accumulation. We also establish an exact extension to non-Gaussian
$SU(1,1)$ sectors. For a lowest-weight Fock seed with conserved number difference $d$, the same ratio equation governs the evolution, while the reduced spectrum becomes a negative-binomial distribution determined by $(q_k,d)$. The Gaussian formulas are recovered at $d=0$. Cosmological perturbations and a chirped-pulse optical parametric amplifier illustrate the common dynamical mechanism and its information-theoretic consequences.

\end{abstract}

\maketitle

\section{introduction}
\label{introduction}
Pair production is a generic consequence of quantum evolution in time-dependent backgrounds and parametrically driven systems. When translational symmetry is preserved, excitations are naturally produced in pairs of opposite momenta, and the corresponding quantum state takes the
form of a two-mode squeezed state. A prominent realization occurs for primordial perturbations generated during inflation, where the modes $\mathbf{k}$ and $-\mathbf{k}$ become strongly correlated as the background
evolves \cite{Grishchuk:1990bj,Albrecht:1992kf,Polarski:1995jg}. This squeezed-state description has played a central role in discussions of the quantum-to-classical transition of cosmological fluctuations \cite{Kiefer:1998jk,Kiefer:2006je,Kiefer:2008ku,Martin:2012pea,Martin:2015qta}, as well as in studies of their entropy, decoherence, and observable quantum correlations \cite{Campo:2008ju,Prokopec:2006fc,Nelson:2016kjm}. The same quadratic pair-production structure also appears in two-photon quantum optics and continuous-variable quantum information \cite{Caves:1985zz,Schumaker:1985zz,Braunstein:2005zz}, where Gaussian-state methods provide a general description of squeezing and mode correlations \cite{Weedbrook:2011wxo,Adesso:2007tx,Adesso:2014npz,Simon:1999lfr,Duan:2000awi}. For a complete two-mode squeezed state, the correlations between the two partner modes are retained and the total state may remain pure. An observer
with access to only one member of the pair instead assigns a reduced density matrix obtained by tracing over the inaccessible partner. The purity and entropies of this reduced state quantify the information hidden in the two-mode correlations. This reduced-state mixedness should be distinguished from genuine decoherence of the complete system caused by an external environment \cite{Feynman:1963fq,Hu:1991di,Weenink:2011dd,breuer2002theory,Campo:2005sy,Burgess:2006jn}. The latter requires nonunitary reduced dynamics and, in its Markovian limit, is described by quantum dynamical semigroups or master equations \cite{Gorini:1975nb,Lindblad:1975ef}. In the present work, the information loss under consideration is primarily the loss associated with tracing over one member
of a correlated pair.

The dynamics of a two-mode squeezed state is conventionally described by a squeezing amplitude $r_k(\eta)$ and a squeezing phase $\phi_k(\eta)$, or equivalently by a pair of Bogoliubov coefficients
$\alpha_k(\eta)$ and $\beta_k(\eta)$. These parametrizations are complete, but their direct implementation can become cumbersome in general time-dependent systems. The variables $r_k$ and $\phi_k$ satisfy coupled nonlinear equations, and the squeezing phase can undergo rapid oscillations. Moreover, when different backgrounds or driving protocols are compared, the two-mode state and its reduced density matrix are often reconstructed separately for each model before the corresponding information-theoretic quantities are evaluated. Additional deformation parameters, such as those appearing in normalized open two-mode squeezed states, further complicate the state-level expressions. Recent studies have investigated the purity, entropy, and squeezing of inflationary perturbations with nontrivial propagation speed \cite{Liu:2021nzx,Liu:2025caj,Liu:2026mzz}, as well as the modifications produced by K-essence dynamics and modified dispersion relations \cite{Li:2021kfq,Li:2023ekd}. Krylov-space techniques provide a complementary description of state and operator growth in early-universe systems \cite{Li:2024kfm,Li:2024ljz,Zhai:2024odw,Zhai:2025abc,Parker:2018yvk,Liu:2022god,Bhattacharya:2023zqt}, and have recently been related to two-mode information geometry and field-theory correlation functions \cite{Zhai:2024tkz,He:2025guu,Liu:2018hno}. These developments motivate a formulation in which the model-dependent part of the evolution is clearly separated from the universal information-theoretic structure of a normalized two-mode pair state. In particular, for the deformed paired states considered here, it is desirable to identify a single closed dynamical variable that both incorporates the state-level deformation and determines the spectrum of the one-mode reduced density matrix.

In this work, we establish such a formulation in terms of the Bogoliubov ratio
\begin{equation}
\lambda_k(\eta)
\equiv
\frac{\beta_k(\eta)}{\alpha_k(\eta)},
\qquad
|\alpha_k|^2-|\beta_k|^2=1,
\qquad
|\lambda_k|<1 .
\label{eq:intro_lambda}
\end{equation}
We build on the normalized open two-mode squeezed state obtained previously and show that, within its convergence domain, it admits the exact state-level representation
\begin{equation}
|\psi_k(\eta)\rangle
=
\sqrt{1-|\lambda_k(\eta)|^2}
\sum_{n=0}^{\infty}
\lambda_k^n(\eta)
|n_k,n_{-k}\rangle .
\label{eq:intro_state}
\end{equation}
Here the term ``open'' refers to the origin of the deformation parameters in the previously constructed state. The Bogoliubov representation established
below is a normalized parametrization of that state and does not, by itself, imply that its original open-system evolution is generated by a Gaussian unitary or by the Hamiltonian dynamics introduced subsequently.

For a general time-dependent quadratic two-mode Hamiltonian,
\begin{equation}
\hat H_k(\eta)
=
\omega_k(\eta)
\left(
\hat a_k^\dagger\hat a_k
+
\hat a_{-k}^\dagger\hat a_{-k}
+1
\right)
+
i g_k(\eta)
\left(
\hat a_k^\dagger\hat a_{-k}^\dagger
-
\hat a_k\hat a_{-k}
\right),
\label{eq:intro_hamiltonian}
\end{equation}
the Schr\"odinger equation reduces to the closed Riccati-type equation
\begin{equation}
\frac{d\lambda_k}{d\eta}
=
g_k(\eta)\left(1-\lambda_k^2\right)
-
2i\omega_k(\eta)\lambda_k .
\label{eq:intro_riccati}
\end{equation}
The complex variable $\lambda_k$ therefore combines the information ordinarily carried by the two real variables $r_k$ and $\phi_k$. The function $g_k$ controls pair creation and annihilation, whereas $\omega_k$ generates the phase rotation of the mode pair. Their competition
determines both the magnitude and phase of the Bogoliubov ratio. After the $-\mathbf{k}$ mode is traced out, the phase of $\lambda_k$ drops out and the reduced-state spectrum becomes
\begin{equation}
p_n(\eta)
=
\left[1-q_k(\eta)\right]q_k^n(\eta),
\qquad
q_k(\eta)
\equiv
|\lambda_k(\eta)|^2 ,
\label{eq:intro_spectrum}
\end{equation}
which is the geometric spectrum associated with a two-mode Gaussian squeezed state \cite{Serafini:2006eim,brask2021gaussian}. The purity, linear entropy, R\'enyi-2 entropy, and
von Neumann entropy are consequently determined by the same real quantity $q_k$ through the standard definitions \cite{renyi1961measures,Manfredi:2000qzn,nielsen2010quantum}. The complete calculation can thus be organized as
\begin{equation}
\bigl\{\omega_k(\eta),g_k(\eta)\bigr\}
\longrightarrow
\lambda_k(\eta)
\longrightarrow
q_k(\eta)
\longrightarrow
\bigl\{
\mu_k,\,
S_{L,k},\,
S_{2,k},\,
S_{{\rm vN},k}
\bigr\}.
\label{eq:intro_map}
\end{equation}
We emphasize that the map in Eq.~\eqref{eq:intro_map} is an exact parametrization of the vacuum Gaussian pair sector rather than a new definition of quantum-information measures. For this sector, the same purity and entropies can be obtained from the	reduced density matrix or from the symplectic eigenvalue of the covariance matrix. The utility of the Bogoliubov-ratio representation is instead dynamical: it separates the model-dependent evolution of a single complex group coordinate from the subsequent evaluation of the reduced-state spectrum. Moreover, the ratio dynamics is not restricted to the Gaussian vacuum representation. The same $SU(1,1)$ coordinate extends to non-Gaussian squeezed Fock sectors when the corresponding representation label is retained. We make both the equivalence and this extension explicit below, and we also state where a scalar ratio ceases to be a complete state descriptor.

The first step contains the model-dependent dynamics, whereas the conversion from $q_k$ to the reduced-state diagnostics is universal for the normalized
paired state. This separation avoids reconstructing and diagonalizing the reduced density matrix independently for every realization of the time-dependent quadratic Hamiltonian.

We apply the framework first to cosmological perturbations. In this case, the evolving background determines the effective frequency and the pair-production coupling, thereby changing the competition between phase rotation and squeezing. The resulting trajectory of $\lambda_k$ directly determines the growth of the pair correlations, the reduction of one-mode purity, and the increase of the reduced-state entropies. This construction is closely related to the standard Bogoliubov description of particle creation in expanding backgrounds \cite{parker1969quantized,parker1971quantized,birrell1984quantum} and to its systematic formulation in quantum field theory in curved spacetime \cite{fulling1989aspects,wald1994quantum,parker2009quantum}. Throughout the calculation, the reduced-state diagnostics are evaluated only within the normalized Bogoliubov domain $|\lambda_k|<1$. As an independent realization and an analytic consistency test, we also consider a nondegenerate optical parametric amplifier. In the resonant
constant-coupling limit, Eq.~\eqref{eq:intro_riccati} gives
$\lambda_{\rm opt}=\tanh\tau$, reproducing the standard two-mode squeezing solution. For a finite Gaussian pump with a time-dependent detuning, the production of correlations is localized near the pulse center, while the chirp suppresses their subsequent accumulation. The corresponding reduced-state quantities therefore exhibit a delayed response and approach finite late-time values. Such finite-duration pair-production processes are supported by pulsed squeezing experiments and multimode optical analyses \cite{Slusher:1985zz,yurke1987generating,wasilewski2006pulsed,christ2012probing}, while chirped and quasi-phase-matched parametric amplification provides the relevant time-dependent driving protocols \cite{brecht2015photon,Lvovsky:2006vpf,Horoshko:2013zdl,cui2020high}. The cosmological and optical systems have different microscopic origins, but both are governed by the same quadratic pairing algebra and enter the
present construction only through the functions $\omega_k$ and $g_k$. The Bogoliubov-ratio formulation therefore isolates the universal information content of two-mode pair production from the physical mechanism responsible for the evolution. The same structural correspondence also appears in the dynamical Casimir effect \cite{Moore:1970tmc,Dodonov:2010zza,nation2012colloquium} and in superconducting-circuit realizations of parametrically generated photon pairs \cite{Johansson:2009zz,Wilson:2011rsw}. The present framework may consequently be extended to other time-dependent quadratic bosonic systems, provided that the evolved state retains a normalized paired Bogoliubov representation.

The remainder of this paper is organized as follows. In
Sec.~\ref{sec:two-mode-hamiltonian}, we will introduce the general time-dependent two-mode Hamiltonian. In Secs.~\ref{wave function1} and \ref{sec:bogoliubov-representation}, we will review the normalized deformed two-mode state and establish its Bogoliubov representation. In Sec.~\ref{purity abd entropy}, we will derive the evolution equation for $\lambda_k$, the reduced-state spectrum, and the associated information-theoretic quantities. The cosmological and optical realizations, together with their numerical solutions, will be presented in Sec. \ref{sec:numerical-solutions}. We will summarize the results and discuss possible
extensions in Sec. \ref{sec:conclusion-outlook}, while the detailed derivations are collected in the appendices.

\section{Hamiltonian of the two-mode squeezed state}
\label{sec:two-mode-hamiltonian}

We begin with a Hamiltonian that is independent of any particular cosmological background or optical driving protocol. Both the free-mode frequency and the squeezing coupling may vary with time. The resulting quadratic form describes pair production in terms of two model-dependent functions.

We consider two bosonic modes with opposite momenta, labelled by $k$ and $-k$. Their annihilation and creation operators satisfy
\begin{equation}
[\hat a_k,\hat a_k^\dagger]=1,
\qquad
[\hat a_{-k},\hat a_{-k}^\dagger]=1,
\label{eq:canonical-commutators}
\end{equation}
with all commutators between independent modes vanishing. Spatial translation symmetry enforces conservation of total momentum, so pair excitations are naturally described in the $(k,-k)$ basis. The corresponding bilinear generators produce $\mathrm{SU}(1,1)$ transformations, which underlie the standard construction and classification of two-mode squeeze operators \cite{truax1985baker,Stephan:2009pn}. The same paired structure appears in two-photon quantum optics and in squeezed-state descriptions of primordial perturbations and relic gravitons \cite{Grishchuk:1990bj,Albrecht:1992kf,Caves:1985zz,Schumaker:1985zz}.

The number-conserving contribution to the Hamiltonian is
\begin{equation}
\hat H_{0,k}(\eta)
=
\omega_k(\eta)
\left(
\hat a_k^\dagger \hat a_k
+
\hat a_{-k}^\dagger \hat a_{-k}
+1
\right),
\label{eq:free-two-mode-hamiltonian}
\end{equation}
where $\eta$ is the evolution parameter and $\omega_k(\eta)$ is the effective frequency of the mode pair. A constant choice, such as $\omega_k=k$, gives the usual fixed-frequency phase rotation. Allowing $\omega_k$ to depend on time accommodates the effective frequencies generated by an evolving background, a time-dependent mass, or a modified dispersion relation. Such time-dependent frequencies are standard in Bogoliubov descriptions of particle production in nonstationary and curved spacetimes \cite{parker1969quantized,parker1971quantized,birrell1984quantum,parker2009quantum,higuchi2009conformally}. In the present framework, this freedom generalizes the free evolution without altering the two-mode operator algebra.

The pair-production contribution is taken to be
\begin{equation}
\hat H_{\mathrm{sq},k}(\eta)
=
i g_k(\eta)
\left(
\hat a_k^\dagger \hat a_{-k}^\dagger
-
\hat a_k \hat a_{-k}
\right).
\label{eq:squeezing-two-mode-hamiltonian}
\end{equation}
The creation term generates a pair of excitations with opposite momenta, whereas the annihilation term removes such a pair. For real $g_k(\eta)$, their relative sign and the factor of $i$ make Eq.~\eqref{eq:squeezing-two-mode-hamiltonian} Hermitian. A possible phase of a complex pairing amplitude can equivalently be transferred to the mode operators by a phase convention, so the real-coupling form is sufficient for the models considered below. This interaction is the standard source of two-mode squeezing and pair correlations \cite{Caves:1985zz,Schumaker:1985zz,dodonov2002nonclassical,walls2008quantum}.

The function $g_k(\eta)$ measures the instantaneous strength of pair production. In a parametric amplifier it is fixed by the nonlinear interaction and the pump profile, while in cosmological applications it is induced by the time-dependent background and governs the amplification of vacuum fluctuations \cite{Grishchuk:1990bj,Albrecht:1992kf,Polarski:1995jg}. These realizations have different microscopic origins but share the same quadratic pair-production structure, a correspondence that has long connected cosmological particle creation with laboratory squeezing \cite{Grishchuk:1992tw}. We therefore leave $g_k(\eta)$ unspecified until a particular model is introduced.

Combining Eqs.~\eqref{eq:free-two-mode-hamiltonian} and \eqref{eq:squeezing-two-mode-hamiltonian}, the Hamiltonian of the mode pair is
\begin{equation}
\boxed{
	\hat H_k(\eta)
	=
	\omega_k(\eta)
	\left(
	\hat a_k^\dagger \hat a_k
	+
	\hat a_{-k}^\dagger \hat a_{-k}
	+1
	\right)
	+
	i g_k(\eta)
	\left(
	\hat a_k^\dagger \hat a_{-k}^\dagger
	-
	\hat a_k \hat a_{-k}
	\right)
}
\label{eq:general-two-mode-hamiltonian}
\end{equation}
with real functions $\omega_k(\eta)$ and $g_k(\eta)$. The former controls the free phase rotation, whereas the latter controls transitions between adjacent paired number states. Their competition determines both the magnitude and phase of the resulting squeezing. Because the operator content of Eq.~\eqref{eq:general-two-mode-hamiltonian} remains fixed, different cosmological and optical models can be implemented by changing only these two input functions.

Under Eq.~\eqref{eq:general-two-mode-hamiltonian}, quadratic evolution maps the initial mode operators linearly into creation and annihilation operators while preserving the canonical commutators. If the resulting state admits a normalized paired Bogoliubov representation, its dynamics can be expressed through the ratio $\lambda_k=\beta_k/\alpha_k$. The functions $\omega_k(\eta)$ and $g_k(\eta)$ determine its trajectory, and the reduced-state purity and entropies follow from $|\lambda_k|^2$ through the relations derived below.

\section{Wave function} 
\label{wave function1}

We use the generalized two-mode wave function derived from the Meixner-polynomial construction of Ref.~\cite{Zhai:2024tkz}. Here it is employed as a parametrization of a Hermitian two-mode state rather than as the state of an open quantum system. In the Krylov representation, the $n$-th basis vector is identified with the paired two-mode number state,
\begin{equation}
|e_n) \equiv |n,n\rangle_{\vec k,-\vec k},
\qquad
|e_0) \equiv |0,0\rangle_{\vec k,-\vec k}.
\label{eq:krylov-paired-basis}
\end{equation}
This identification is natural for a two-mode squeezed system, because the elementary excitations are created in correlated pairs with momenta $\vec k$ and $-\vec k$. The initial Krylov vector is mapped to the two-mode vacuum, while the higher Krylov vectors describe paired number states. This is the same basis choice used in the generalized two-mode construction of Ref.~\cite{Zhai:2024tkz}. The underlying pair generators produce $\mathrm{SU}(1,1)$ transformations, whose symplectic representation provides the group-theoretic basis for two-mode squeezing, Bogoliubov transformations, and the paired number-state expansion \cite{truax1985baker,Stephan:2009pn,dodonov2002nonclassical,Marian:1993zz}.

The generalized Lanczos recursion contains an additional diagonal contribution and takes the form
\begin{equation}
\mathcal{L}|O_n)
=
-i c_n |O_n)
+
b_{n+1}|O_{n+1})
+
b_n|O_{n-1}) ,
\label{eq:generalized-lanczos-recursion}
\end{equation}
where $b_n$ describes nearest-neighbor hopping along the Krylov chain and $c_n$ specifies the diagonal contribution. The class of models considered in Ref.~\cite{Zhai:2024tkz} uses
\begin{equation}
b_n^2
=
|1-u_1^2|n(n-1+\chi),
\qquad
c_n
=
i u_2(2n+\chi).
\label{eq:lanczos-coefficients}
\end{equation}
The parameters $u_1$ and $u_2$ are not required to be constants. For a time-dependent quadratic Hamiltonian, they may inherit the evolution-parameter dependence of the instantaneous Lanczos coefficients, and we write $u_j=u_j(\eta)$ when this dependence needs to be displayed explicitly. 
The two-mode Hamiltonian conserves the number difference
	\begin{equation}
	\hat D=
	\hat a_k^\dagger\hat a_k
	-\hat a_{-k}^\dagger\hat a_{-k},
	\qquad
	[\hat H_k,\hat D]=0.
	\end{equation}
	In the sector $D=d\geq0$, where $d$ is a nonnegative integer, a convenient basis is
	\begin{equation}
	|e_n^{(d)}\rangle
	=
	|n+d,n\rangle_{k,-k}.
	\end{equation}
	The corresponding $SU(1,1)$ ladder matrix elements give
	\begin{equation}
	b_n(\eta)=|g_k(\eta)|\sqrt{n(n+d)},
	\end{equation}
	whereas the diagonal contribution is proportional to $2n+d+1$. Consequently, the Meixner parameter in Eq.~\eqref{eq:lanczos-coefficients} is
\begin{equation}
	\chi=d+1.
	\end{equation}
	The vacuum pair sector used in the cosmological and optical 	applications corresponds to $d=0$, for which $\chi=1$ and $b_n=|g_k|n$. In the remainder of the present Gaussian
	derivation we set $d=0$. The sectors with $d\geq1$ provide an exact non-Gaussian extension and will be considered in Sec.~\ref{sec:nonGaussian_extension}. Specializing to the vacuum pair sector $d=0$, one obtains
\begin{equation*}
b_n(\eta)=|g_k(\eta)|\,n,
\qquad
-i c_n(\eta)=\omega_k(\eta)(2n+1),
\end{equation*}
and therefore
\begin{equation*}
\boxed{
	u_2(\eta)=\omega_k(\eta),
	\qquad
	\left|1-u_1^2(\eta)\right|=|g_k(\eta)|^2
}
\end{equation*}
in the dimensionless normalization adopted for the recurrence. If the Hamiltonian is instead normalized by a fixed reference frequency $\Omega_0$, the corresponding relations are $u_2=\omega_k/\Omega_0$ and $|1-u_1^2|=|g_k/\Omega_0|^2$. Thus $u_2$ parametrizes the diagonal sector, while $\sqrt{|1-u_1^2|}$ gives the magnitude of the off-diagonal pair-production coupling; the sign or phase of $g_k$ is carried by the squeezing phase. This is a coefficient-level correspondence in the Krylov tridiagonal representation. In the present Hermitian system, $u_2$ therefore represents the free phase-rotation sector rather than environmental dissipation.

A concrete cosmological realization makes the time dependence explicit. For a modified dispersion function $f(\eta)$ and scale factor $a(\eta)$, one obtains
\begin{equation*}
u_2(\eta)
=
\frac{k}{2}\left[f^2(\eta)+1\right]
=
\omega_k(\eta),
\end{equation*}
and
\begin{equation*}
\left|1-u_1^2(\eta)\right|
=
\frac{k^2}{4}\left[f^2(\eta)-1\right]^2
+
\left[\frac{a'(\eta)}{a(\eta)}\right]^2
=
|g_k(\eta)|^2.
\end{equation*}
Hence both deformation parameters generally evolve with the background rather than remaining fixed phenomenological constants. Their dependence on the scale factor, modified dispersion relations, effective masses, and different cosmological epochs has been investigated systematically in Refs.~\cite{Li:2024kfm,Zhai:2024odw,Zhai:2025abc}, while the model-independent Meixner-polynomial construction is developed in Ref.~\cite{Zhai:2024tkz}. For compactness, the arguments of $u_1(\eta)$ and $u_2(\eta)$ are suppressed in the equations below unless their time dependence is central to the discussion.

The parameter $u_1$ modifies the off-diagonal hopping in Krylov space, whereas $u_2$ enters through the diagonal term and is identified here with the effective frequency $\omega_k$. With $\tilde c_n=-ic_n$, the recurrence can be written as
\begin{equation}
P_{n+1}(x)
=
(x-\tilde c_n)P_n(x)
-
b_n^2P_{n-1}(x).
\label{eq:meixner-recurrence}
\end{equation}
This recurrence defines the Meixner polynomials of the second kind. Their appearance follows directly from the generalized Lanczos coefficients and allows the Krylov expansion to be resummed with the corresponding generating function, as done for the generalized two-mode squeezed state in Ref.~\cite{Zhai:2024tkz}. Relations between tridiagonal Lanczos recursions and orthogonal polynomials are also standard in the recursion method and in the general theory of orthogonal polynomial sequences \cite{viswanath1994recursion,hetyei2010meixner,koekoek1996askey}.

Using the generating function of the Meixner polynomials of the second kind, Ref.~\cite{Zhai:2024tkz} obtains the generalized two-mode squeezed state
\begin{equation}
|O(\eta))
=
\frac{\operatorname{sech} r_k}
{1+u_2\tanh r_k}
\sum_{n=0}^{\infty}
|1-u_1^2|^{n/2}
\left[
-\frac{
	e^{2i\phi_k}\tanh r_k
}
{
	1+u_2\tanh r_k
}
\right]^n
|e_n),
\label{eq:generalized-two-mode-state}
\end{equation}
where $r_k\equiv r_k(\eta)$ and $\phi_k\equiv\phi_k(\eta)$ are the squeezing parameter and squeezing angle. This expression reduces to the ordinary two-mode squeezed state when the deformation is removed. We use its normalized form below.
The resulting normalized wave function is
\begin{equation}
\begin{aligned}
|O(\eta)\rangle
=&
\frac{
	\sqrt{
		(1+u_2\tanh r_k)^2
		-
		|1-u_1^2|\tanh^2 r_k
	}
}
{
	1+u_2\tanh r_k
}
\sum_{n=0}^{\infty}
|1-u_1^2|^{n/2}
\\
&\times
\left[
-\frac{
	e^{2i\phi_k}\tanh r_k
}
{
	1+u_2\tanh r_k
}
\right]^n
|n,n\rangle_{\vec k,-\vec k}.
\end{aligned}
\label{eq:normalized-generalized-two-mode-state}
\end{equation}
The detailed normalization procedure is given in Appendix~A of Ref.~\cite{Liu:2026mzz}. The prefactor in Eq.~\eqref{eq:normalized-generalized-two-mode-state} provides the normalized parametrization used in the following sections. The state retains the paired number basis $|n,n\rangle_{\vec k,-\vec k}$ generated by the standard $\mathrm{SU}(1,1)$ two-mode squeezing algebra \cite{truax1985baker,Stephan:2009pn}, while $u_1$ and $u_2$ deform the relative amplitudes through the generalized Lanczos coefficients. Its geometric amplitude can be reorganized into a normalized Bogoliubov form, allowing the ratio $\lambda_k=\beta_k/\alpha_k$ to be introduced in the next section. In the undeformed limit $u_2\to0$ and $|1-u_1^2|\to1$, the prefactor becomes $\operatorname{sech}r_k$ and Eq.~\eqref{eq:normalized-generalized-two-mode-state} reduces continuously to the conventional two-mode squeezed vacuum \cite{Caves:1985zz,Schumaker:1985zz,dodonov2002nonclassical}.

\section{Bogoliubov representation}
\label{sec:bogoliubov-representation}

The normalized generalized two-mode squeezed state has the paired geometric structure familiar from Gaussian quantum theory. In our work \cite{ma2026krylov}, we have demonstrated that the wave function can be represented by the Bogoliubov transformation when the Hamiltonian is of the group structure. Pure two-mode squeezed states generated by quadratic bosonic Hamiltonians can be represented in the paired number basis or by a canonical Bogoliubov transformation \cite{Caves:1985zz,Schumaker:1985zz,Weedbrook:2011wxo,brask2021gaussian}. In the Meixner-polynomial solution, $u_1$ and $u_2$ change the relative amplitudes but leave the expansion over $|n,n\rangle_{\vec k,-\vec k}$ intact.

The normalized state can be written as
\begin{equation}
|O(\eta)\rangle
=
\sqrt{1-|\lambda_k|^2}
\sum_{n=0}^{\infty}
\lambda_k^n
|n,n\rangle_{\vec k,-\vec k},
\label{eq:generalized-state-lambda-form}
\end{equation}
 Equation~\eqref{eq:generalized-state-lambda-form} represents the $d=0$ vacuum sector of the two-mode $SU(1,1)$ representation. Its geometric Schmidt coefficients identify it as a pure Gaussian two-mode squeezed
state. The non-Gaussian sectors $d\geq1$ retain the same Bogoliubov-ratio dynamics but acquire additional representation-dependent coefficients; they are derived in Sec.~\ref{sec:nonGaussian_extension}. The paired geometric expansion in Eq.~\eqref{eq:generalized-state-lambda-form} has the same number-state structure as the canonical two-mode squeezed vacuum introduced in the standard theory of two-mode squeezing and subsequently used throughout continuous-variable quantum optics \cite{Caves:1985zz,Schumaker:1985zz,Braunstein:2005zz,Weedbrook:2011wxo,walls2008quantum}. The model dependence is compressed into the single complex ratio $\lambda_k$, while the deformation parameters determine how this ratio differs from its standard undeformed value, where
\begin{equation}
\lambda_k
=
-\frac{
	\sqrt{|1-u_1^2|}\,e^{2i\phi_k}\tanh r_k
}
{
	1+u_2\tanh r_k
}.
\label{eq:lambda-generalized-deformation}
\end{equation}
For comparison, a two-mode Bogoliubov vacuum has the number-state expansion
\begin{equation}
|0_k\rangle_{\mathrm{B}}
=
\sqrt{1-\left|\frac{\beta_k}{\alpha_k}\right|^2}
\sum_{n=0}^{\infty}
\left(
\frac{\beta_k}{\alpha_k}
\right)^n
|n,n\rangle_{\vec k,-\vec k},
\label{eq:bogoliubov-vacuum-expansion}
\end{equation}
where the coefficients obey the bosonic normalization condition
\begin{equation}
|\alpha_k|^2-|\beta_k|^2=1.
\label{eq:bogoliubov-normalization}
\end{equation}
Comparing Eqs.~\eqref{eq:generalized-state-lambda-form} and \eqref{eq:bogoliubov-vacuum-expansion} gives the central identification
\begin{equation}
\boxed{
	\lambda_k
	=
	\frac{\beta_k}{\alpha_k}
	=
	-\frac{
		\sqrt{|1-u_1^2|}\,e^{2i\phi_k}\tanh r_k
	}
	{
		1+u_2\tanh r_k
	}
}.
\label{eq:bogoliubov-ratio-identification}
\end{equation}
Thus the two real squeezing variables and the state deformation enter the subsequent calculation through one complex ratio. For real $u_2$ and $1+u_2\tanh r_k>0$, the phase convention $\alpha_k>0$ yields
\begin{equation}
\alpha_k
=
\frac{
	1+u_2\tanh r_k
}
{
	\sqrt{
		(1+u_2\tanh r_k)^2
		-|1-u_1^2|\tanh^2 r_k
	}
},
\label{eq:generalized-alpha-coefficient}
\end{equation}
and
\begin{equation}
\beta_k
=
-\frac{
	\sqrt{|1-u_1^2|}\,e^{2i\phi_k}\tanh r_k
}
{
	\sqrt{
		(1+u_2\tanh r_k)^2
		-|1-u_1^2|\tanh^2 r_k
	}
}.
\label{eq:generalized-beta-coefficient}
\end{equation}
In the undeformed limit $u_1=u_2=0$, Eqs.~\eqref{eq:generalized-alpha-coefficient} and \eqref{eq:generalized-beta-coefficient} reduce to the canonical Bogoliubov coefficients of a two-mode squeezed vacuum,
\begin{equation*}
\alpha_k^{(0)}=\cosh r_k,
\qquad
\beta_k^{(0)}=-e^{2i\phi_k}\sinh r_k,
\qquad
\frac{\beta_k^{(0)}}{\alpha_k^{(0)}}
=-e^{2i\phi_k}\tanh r_k.
\end{equation*}
Accordingly, Eq.~\eqref{eq:generalized-state-lambda-form} becomes
\begin{equation*}
|O(\eta)\rangle
\longrightarrow
\operatorname{sech}r_k
\sum_{n=0}^{\infty}
\left(-e^{2i\phi_k}\tanh r_k\right)^n
|n,n\rangle_{\vec k,-\vec k},
\end{equation*}
which is the standard two-mode squeezed-vacuum state \cite{Caves:1985zz,Schumaker:1985zz,Braunstein:2005zz,Weedbrook:2011wxo}. The minus sign and the factor $2\phi_k$ reflect the phase convention adopted here; they may be shifted by redefining the mode phases and do not alter the occupation statistics or the reduced-state diagnostics. Thus $u_1$ and $u_2$ quantify departures from the canonical closed two-mode squeezed form, whereas setting both parameters to zero restores the familiar unitary squeezing limit. These expressions satisfy Eq.~\eqref{eq:bogoliubov-normalization} and reproduce the normalization factor in Eq.~\eqref{eq:generalized-state-lambda-form}. The rewriting therefore establishes an exact state-level Bogoliubov representation for the Hermitian two-mode dynamics considered here. The relevant group-theoretic structure is the standard $\mathrm{SU}(1,1)$ structure of two-mode squeezing \cite{truax1985baker,Stephan:2009pn,dodonov2002nonclassical}.

The Bogoliubov form combines the two real squeezing variables into a single closed complex coordinate. Cosmological squeezing is commonly described by the coupled amplitude $r_k(\eta)$ and phase $\phi_k(\eta)$. These variables have a clear physical interpretation, but the phase can oscillate rapidly in time-dependent backgrounds. Our recent analysis of cosmological perturbations with nontrivial propagation dynamics evaluates purity and entropic quantities in terms of them \cite{Liu:2026mzz}. Here they are combined into the single complex quantity $\lambda_k$, which obeys a closed Riccati-type equation for a general quadratic two-mode Hamiltonian.

After tracing over the partner mode, the phase of $\lambda_k$ drops out. Defining
\begin{equation}
q_k
=
\left|\frac{\beta_k}{\alpha_k}\right|^2
=
|\lambda_k|^2,
\label{eq:q-from-bogoliubov-ratio}
\end{equation}
the one-mode reduced density matrix has the geometric spectrum
\begin{equation}
p_n=(1-q_k)q_k^n.
\label{eq:reduced-geometric-spectrum-bogoliubov}
\end{equation}
This spectrum is the standard reduced spectrum of a pure two-mode squeezed Gaussian state and is fully determined by the reduced symplectic eigenvalue \cite{Adesso:2014npz,serafini2004symplectic,demarie2012pedagogical,eisert2003introduction,horodecki2009quantum}. Consequently, the purity and entropies can be evaluated directly from $q_k$, without reconstructing the reduced density matrix separately for each model. The resulting loss of one-mode purity reflects the correlations hidden by the trace over the $-k$ mode; it should not be identified with decoherence of the complete two-mode state caused by an explicit external environment. The Bogoliubov ratio therefore supplies the link between the normalized wave function and the unified reduced-state diagnostics developed below.

\section{Purity and entropy from the $\lambda_k$ framework}
\label{purity abd entropy}
For a normalized two-mode wave function in Bogoliubov form, the reduced-state diagnostics can be obtained from the complex ratio
\begin{equation}
\lambda_k=\frac{\beta_k}{\alpha_k},
\qquad
q_k=|\lambda_k|^2,
\label{eq:lambda-bogoliubov-ratio}
\end{equation}
where the Bogoliubov coefficients satisfy $|\alpha_k|^2-|\beta_k|^2=1$. After the $-k$ mode is traced out, the phase of $\lambda_k$ drops out and the reduced spectrum depends only on $q_k$. The purity, linear entropy, R\'enyi-2 entropy, and von Neumann entropy can therefore be written directly as
\begin{equation}
\begin{aligned}
\mu_k
&=
\frac{1-|\lambda_k|^2}{1+|\lambda_k|^2},
&
S_{L,k}(\eta)
&=
\frac{2|\lambda_k|^2}{1+|\lambda_k|^2},
\\
S_{2,k}
&=
\ln\!\left[
\frac{1+|\lambda_k|^2}{1-|\lambda_k|^2}
\right],
&
S_{\mathrm{vN},k}
&=
-\ln\!\left[1-|\lambda_k|^2\right]
-\frac{|\lambda_k|^2}{1-|\lambda_k|^2}
\ln|\lambda_k|^2.
\end{aligned}
\label{eq:lambda-general-diagnostics}
\end{equation}
The information measures depend on the model only through the Bogoliubov ratio. In the vacuum limit, $\lambda_k=0$, the purity is $\mu_k=1$ and all entropies vanish. As $|\lambda_k|$ increases, stronger correlations between the $k$ and $-k$ modes reduce the one-mode purity and raise the entropies. Directly evolving $\lambda_k$ also removes the need to use $r_k$ and $\phi_k$ as intermediate variables.

The functions $\omega_k$ and $g_k$ may be any prescribed functions of time, provided that the evolution remains compatible with a normalized Bogoliubov two-mode state. Once they specify a physical model, the same evolution equation determines its Bogoliubov ratio, and Eq.~\eqref{eq:lambda-general-diagnostics} gives the purity and entropies without another reduced-density-matrix calculation. Sections~5.2 and 5.3 apply this construction to cosmological perturbations and a time-dependent optical parametric amplifier. The procedure also applies to other quadratic bosonic systems governed by two-mode pair production.

\subsection{Evolution equation for $\lambda_k$}

The normalized two-mode state is written as
\begin{equation}
|\psi_k(\eta)\rangle
=
\sqrt{1-|\lambda_k(\eta)|^2}
\sum_{n=0}^{\infty}
\lambda_k^n(\eta)|n_k,n_{-k}\rangle,
\label{eq:lambda-two-mode-state}
\end{equation}
This form is equivalent to the Bogoliubov representation of a two-mode squeezed Gaussian state \cite{Caves:1985zz,Schumaker:1985zz,brask2021gaussian}. We consider the general quadratic Hamiltonian
\begin{equation}
\hat H_k(\eta)
=
\omega_k(\eta)
\left(
\hat a_k^\dagger\hat a_k
+
\hat a_{-k}^\dagger\hat a_{-k}
+1
\right)
+
i g_k(\eta)
\left(
\hat a_k^\dagger\hat a_{-k}^\dagger
-
\hat a_k\hat a_{-k}
\right),
\label{eq:general-two-mode-hamiltonian-lambda}
\end{equation}
where $\omega_k$ governs the free phase evolution and $g_k$ controls pair creation and annihilation. In units with $\hbar=1$, the state evolves according to the Schr\"odinger equation
\begin{equation}
i\frac{\partial}{\partial\eta}|\psi_k(\eta)\rangle
=
\hat H_k(\eta)|\psi_k(\eta)\rangle.
\label{eq:schrodinger-equation-lambda}
\end{equation}
Substituting Eqs.~\eqref{eq:lambda-two-mode-state} and \eqref{eq:general-two-mode-hamiltonian-lambda} into Eq.~\eqref{eq:schrodinger-equation-lambda}, and then matching the coefficients of each paired number state $|n_k,n_{-k}\rangle$, gives
\begin{equation}
\frac{d\lambda_k}{d\eta}
=
g_k(\eta)\left(1-\lambda_k^2\right)
-2i\omega_k(\eta)\lambda_k.
\label{eq:lambda-riccati-general}
\end{equation}
Equation~\eqref{eq:lambda-riccati-general} contains the complete dynamics required below. Once it is solved, the reduced-state spectrum follows from
\begin{equation}
q_k(\eta)=|\lambda_k(\eta)|^2.
\label{eq:q-lambda-definition}
\end{equation}
The projection leading to Eq.~\eqref{eq:lambda-riccati-general}, together with the construction of the reduced density matrix, is presented in Appendix~\ref{app:lambda-framework}.

	\subsection{Relation to standard Gaussian methods and physical content}
	\label{sec:Gaussian_equivalence}
	For the Gaussian state in Eq.~(28), write
	\begin{equation}
	\lambda_k=\sqrt{q_k}\,e^{i\theta_k}.
	\end{equation}
	The mean occupation number of either member of the pair is
	\begin{equation}
	\bar n_k=\langle\hat a_k^\dagger\hat a_k\rangle	=	\frac{q_k}{1-q_k}.
	\end{equation}
	Using the convention in which the vacuum covariance matrix is
	$\mathbb{I}_2/2$, the symplectic eigenvalue of the one-mode
	reduced covariance matrix is
	\begin{equation}
	\nu_k	=\bar n_k+\frac{1}{2}=\frac{1+q_k}{2(1-q_k)}.
	\end{equation}
	Consequently,
	\begin{equation}
	\mu_k=\frac{1}{2\nu_k}	=\frac{1-q_k}{1+q_k},\qquad
  S_{2,k}=	\ln(2\nu_k)=\ln\frac{1+q_k}{1-q_k},
  \label{muk}
	\end{equation}
	and
	\begin{equation}
	S_{\mathrm{vN},k}	=
	(\bar n_k+1)\ln(\bar n_k+1)
	-\bar n_k\ln\bar n_k.
	\end{equation}
	These equations demonstrate explicitly that, in the Gaussian sector, the Bogoliubov-ratio, reduced-density-matrix, and covariance-matrix descriptions contain the same information.
	The four quantities in Eq.~\eqref{eq:lambda-general-diagnostics} are not independent invariants: $S_{L,k}=1-\mu_k$ and $S_{2,k}=-\ln\mu_k$, while $S_{\mathrm{vN},k}$ is also a function of the same $q_k$. The dynamical interpretation becomes transparent by substituting $\lambda_k=\sqrt{q_k}e^{i\theta_k}$ into Eq.~\eqref{eq:lambda-riccati-general}. For $q_k>0$, one obtains
	\begin{align}
	\frac{dq_k}{d\eta}
	&=
	2g_k\sqrt{q_k}(1-q_k)\cos\theta_k,
	\\
	\frac{d\theta_k}{d\eta}
	&=
	-2\omega_k
	-g_k\frac{1+q_k}{\sqrt{q_k}}\sin\theta_k.
	\end{align}
	At $q_k=0$, the original nonsingular complex equation, Eq.~\eqref{eq:lambda-riccati-general}, should be used. These equations show that $g_k$ supplies the radial	pair-production drive, whereas $\omega_k$ rotates the pair phase. The frequency term therefore affects $q_k$ indirectly by changing the phase alignment entering $\cos\theta_k$.	The ratio does not reveal a new Gaussian invariant, but it makes the competition between resonant pair accumulation and
	phase-mismatch suppression explicit in a common coordinate for the cosmological and optical systems.

	\subsection{Exact non-Gaussian $SU(1,1)$ extension}
	\label{sec:nonGaussian_extension}
	
	The scalar ratio is a complete state coordinate for the vacuum-evolved pure Gaussian pair state. Nevertheless, the same ratio dynamics also acts on non-Gaussian representations	of the two-mode $SU(1,1)$ algebra via \cite{Gerry:91,Truax:1985vk}. Define
	\begin{equation}
	K_+=\hat a_k^\dagger\hat a_{-k}^\dagger,
	\qquad
	K_-=\hat a_k\hat a_{-k},
	\qquad
	K_0=
	\frac{1}{2}
	\left(
	\hat a_k^\dagger\hat a_k
	+\hat a_{-k}^\dagger\hat a_{-k}+1
	\right).
	\end{equation}
	The Hamiltonian in Eq.~\eqref{eq:general-two-mode-hamiltonian-lambda} can then be written as
	\begin{equation}
	\hat H_k
	=
	2\omega_k K_0
	+ig_k(K_+-K_-).
	\end{equation}
	Because the number difference $\hat D$ commutes with all three generators, the evolution separates into invariant sectors labelled by $d$. For $d\geq0$, the state $|d,0\rangle$ is a
	lowest-weight vector,
	\begin{equation}
	K_-|d,0\rangle=0,
	\qquad
	K_0|d,0\rangle
	=
	\frac{d+1}{2}|d,0\rangle.
	\end{equation}
	Using the standard $SU(1,1)$ disentangling of the evolution operator, its action on the lowest-weight state can be written, up to an irrelevant global phase, as
	\begin{equation}
	|\psi_{k,d}(\eta)\rangle
	=
	\left[1-q_k(\eta)\right]^{(d+1)/2}
	\exp\left[\lambda_k(\eta)K_+\right]|d,0\rangle,
	\qquad
	q_k=|\lambda_k|^2<1.
	\end{equation}
	Expanding the exponential gives
	\begin{equation}
	|\psi_{k,d}(\eta)\rangle
	=
	(1-q_k)^{(d+1)/2}
	\sum_{n=0}^{\infty}
	\sqrt{\binom{n+d}{d}}\,
	\lambda_k^n
	|n+d,n\rangle_{k,-k}.
	\label{eq:nonGaussian_state}
	\end{equation}
	For $d=0$, Eq.~\eqref{eq:nonGaussian_state} reduces exactly to the Gaussian state in Eq.~(28). For every $d\geq1$, it is a two-mode squeezed Fock state and is non-Gaussian. A covariance matrix is therefore not, by itself, a complete representation of the state unless the non-Gaussian family and its label $d$ are specified.	Importantly, $\lambda_k$ obeys the same representation-independent Riccati equation,
	\begin{equation}
	\frac{d\lambda_k}{d\eta}
	=
	g_k(1-\lambda_k^2)-2i\omega_k\lambda_k.
	\label{eq:nonGaussian_Riccati}
	\end{equation}
	This result can be verified directly. Writing
	\begin{equation}
	|\psi_{k,d}\rangle
	=
	\sum_{n=0}^{\infty}
	c_n^{(d)}|n+d,n\rangle,
	\end{equation}
	the Schr\"odinger equation gives
	\begin{align}
	i\dot c_n^{(d)}
	={}&
	\omega_k(2n+d+1)c_n^{(d)}
	+ig_k\sqrt{n(n+d)}\,c_{n-1}^{(d)}
	\nonumber\\
	&-ig_k\sqrt{(n+1)(n+d+1)}\,c_{n+1}^{(d)}.
	\label{eq:nonGaussian_recurrence}
	\end{align}
	For the coefficients in Eq.~\eqref{eq:nonGaussian_state},
	\begin{equation}
	\frac{
		\sqrt{n(n+d)}\,c_{n-1}^{(d)}}{c_n^{(d)}}=\frac{n}{\lambda_k},\qquad
\frac{
	\sqrt{(n+1)(n+d+1)}\,c_{n+1}^{(d)}}{c_n^{(d)}}=(n+d+1)\lambda_k.
\end{equation}
Matching the terms proportional to $n$ in Eq.~\eqref{eq:nonGaussian_recurrence} reproduces
Eq.~\eqref{eq:nonGaussian_Riccati}. The remaining $n$-independent equation fixes only the normalization and overall phase. Tracing over the $-k$ mode gives
\begin{equation}
\rho_{k,d}
=
\sum_{n=0}^{\infty}
p_n^{(d)}
|n+d\rangle\langle n+d|,
\end{equation}
where we have defined $p_n^{(d)}=(1-q_k)^{d+1}\binom{n+d}{d} q_k^n$ and it is a normalized negative-binomial distribution because
\begin{equation}
\sum_{n=0}^{\infty} \binom{n+d}{d}q^n=(1-q)^{-(d+1)}.
\end{equation}
The reduced-state purity therefore has the exact closed form
\begin{align}
\mu_{k,d}=\sum_{n=0}^{\infty}
\left[p_n^{(d)}\right]^2=
(1-q_k)^{2(d+1)}
{}_2F_1
\left(
d+1,d+1;1;q_k^2
\right).
\label{eq:nonGaussian_purity}
\end{align}
The linear and R\'enyi-2 entropies are
\begin{equation}
S_{L,k}^{(d)}=1-\mu_{k,d}, \qquad S_{2,k}^{(d)}=
-\ln\mu_{k,d},
\end{equation}
while the von Neumann entropy is
\begin{equation}
S_{\mathrm{vN},k}^{(d)}=-\sum_{n=0}^{\infty} p_n^{(d)} \ln p_n^{(d)}.
\label{eq:nonGaussian_vN}
\end{equation}
For $d=0$, we could uitilize ${}_2F_1(1,1;1;q_k^2)=\frac{1}{1-q_k^2}$ to recover \eqref{muk} defined as $\mu_{k,0}$, and all the Gaussian expressions in Eq.~\eqref{eq:lambda-general-diagnostics} follow. As a simple non-Gaussian example, the $d=1$ sector has
\begin{equation}
p_n^{(1)}=(1-q_k)^2(n+1)q_k^n, ~~\mu_{k,1}
=
\frac{(1-q_k)(1+q_k^2)}{(1+q_k)^3}.
\end{equation}
The extension also specifies the limitation of the scalar framework. For the structured family above, the pair $(\lambda_k,d)$ completely determines the state and its reduced spectrum. For a generic non-Gaussian state
\begin{equation}
|\Psi\rangle
=
\sum_{m,n}c_{mn}|m,n\rangle,
\end{equation}
the single ratio $\lambda_k$ is not a complete state coordinate; the independent amplitudes or higher-order correlations must also be evolved. Thus, the present method extends exactly to non-Gaussian $SU(1,1)$ lowest-weight sectors, but it does not encode arbitrary non-Gaussian states through $\lambda_k$ alone.

\subsection{Cosmological perturbation model}

We apply the $\lambda_k$ description to cosmological perturbations, for which the two-mode squeezed-state formalism describes the correlated modes with comoving momenta $\boldsymbol{k}$ and $-\boldsymbol{k}$ \cite{Grishchuk:1990bj,Albrecht:1992kf,Polarski:1995jg}. In the parametrization adopted here, the effective frequency and squeezing coupling are
\begin{equation}
\omega_k(\eta)
=
\sqrt{k^2+\frac{\alpha}{\eta^2}},
\qquad
g_k(\eta)=-\frac{1}{\eta}.
\label{eq:cosmological-frequency-coupling}
\end{equation}
Here $k=|\boldsymbol{k}|$ is the comoving wavenumber and labels the spatial scale of the perturbation. The dimensionless parameter $\alpha$ controls the background-induced inverse-square contribution to the effective frequency. This structure is the shifted-time form of the standard Mukhanov--Sasaki frequency $k^2-z''/z$ \cite{Mukhanov:1990me,baumann2011physics}. In particular, when $z''/z=(\nu^2-1/4)/\tau^2$ and $\tau=\eta$, the present convention corresponds to $\alpha=1/4-\nu^2$. Thus $\alpha$ encodes the dependence on the cosmological background and the effective mass of the perturbation rather than introducing an additional momentum scale. The coupling $g_k=-1/\eta$ represents the conformal expansion rate in the de Sitter limit, up to the sign convention used in Eq.~\eqref{eq:general-two-mode-hamiltonian-lambda}.

Substitution into Eq.~\eqref{eq:lambda-riccati-general} gives
\begin{equation}
\frac{d\lambda_k}{d\eta}
=
-\frac{1}{\eta}\left(1-\lambda_k^2\right)
-2i\sqrt{k^2+\frac{\alpha}{\eta^2}}\,\lambda_k,
\qquad
\lambda_k(\eta_0)=0.
\label{eq:lambda-cosmological-evolution}
\end{equation}
Defining $q_k^{\mathrm{cos}}(\eta)=|\lambda_k(\eta)|^2$, the reduced density matrix is
\begin{equation}
\rho_k^{\mathrm{cos}}(\eta)
=
\left[1-q_k^{\mathrm{cos}}(\eta)\right]
\sum_{n=0}^{\infty}
\left[q_k^{\mathrm{cos}}(\eta)\right]^n
|n_k\rangle\langle n_k|.
\label{eq:cosmological-reduced-density-matrix}
\end{equation}
The purity, linear entropy, R\'enyi-2 entropy, and von Neumann entropy are defined by
\begin{equation}
\begin{aligned}
\mu_k^{\mathrm{cos}}
&=\operatorname{Tr}\!\left[(\rho_k^{\mathrm{cos}})^2\right],
&
S_{L,k}^{\mathrm{cos}}
&=1-\mu_k^{\mathrm{cos}},
\\
S_{2,k}^{\mathrm{cos}}
&=-\ln\operatorname{Tr}\!\left[(\rho_k^{\mathrm{cos}})^2\right],
&
S_{\mathrm{vN},k}^{\mathrm{cos}}
&=-\operatorname{Tr}\!\left(\rho_k^{\mathrm{cos}}\ln\rho_k^{\mathrm{cos}}\right).
\end{aligned}
\label{eq:cosmological-diagnostic-definitions}
\end{equation}
Their final expressions are
\begin{equation}
\begin{aligned}
\mu_k^{\mathrm{cos}}(\eta)
&=
\frac{1-|\lambda_k(\eta)|^2}{1+|\lambda_k(\eta)|^2},
&
S_{L,k}^{\mathrm{cos}}(\eta)
&=
\frac{2|\lambda_k(\eta)|^2}{1+|\lambda_k(\eta)|^2},
\\
S_{2,k}^{\mathrm{cos}}(\eta)
&=
\ln\!\left[
\frac{1+|\lambda_k(\eta)|^2}{1-|\lambda_k(\eta)|^2}
\right],
&
S_{\mathrm{vN},k}^{\mathrm{cos}}(\eta)
&=
-\ln\!\left[1-|\lambda_k(\eta)|^2\right]
-\frac{|\lambda_k(\eta)|^2}{1-|\lambda_k(\eta)|^2}
\ln|\lambda_k(\eta)|^2.
\end{aligned}
\label{eq:cosmological-diagnostic-results}
\end{equation}
The derivation of Eq.~\eqref{eq:cosmological-diagnostic-results} from the spectrum of Eq.~\eqref{eq:cosmological-reduced-density-matrix} is given in Appendix~\ref{app:lambda-framework}. This cosmological construction provides the first concrete implementation of the unified $\lambda_k$ framework. Different values of $\nu$ generate different effective-frequency evolutions and therefore modify the competition between free phase rotation and two-mode squeezing. To isolate the role of the cosmological frequency term, Sec.~\ref{subsec:cosmological-numerical-solutions} also compares these backgrounds with an ideal squeezing reference in which only the pair-production drive is retained. The parameter choices, initial conditions, and numerical comparisons are deferred to Sec.~\ref{subsec:cosmological-numerical-solutions}.

	In this section, we have shown that $d>0$ describe seeded two-mode parametric processes and are directly applicable to non-vacuum optical inputs. In a cosmological application, the homogeneous vacuum sector considered in the present numerical examples is $d=0$. A nonzero $d$ represents a specified non-vacuum excitation sector and should not be identified directly with
	the standard homogeneous Bunch--Davies vacuum or with primordial non-Gaussianity defined through a curvature bispectrum.

\subsection{Optical two-mode parametric amplifier}

As a non-cosmological realization, we consider a nondegenerate optical parametric amplifier, which generates correlated photon pairs and implements the same quadratic two-mode algebra \cite{Caves:1985zz,Schumaker:1985zz,walls2008quantum}. Its Hamiltonian is
\begin{equation}
\hat H(t)
=
\hbar\omega
\left(
\hat a^\dagger\hat a
+
\hat b^\dagger\hat b
\right)
+
i\hbar\gamma(t)
\left(
\hat a^\dagger\hat b^\dagger
-
\hat a\hat b
\right).
\label{eq:optical-parametric-hamiltonian}
\end{equation}
Here $\omega$ is the carrier angular frequency of the two optical modes in the degenerate-frequency notation, while $\gamma(t)$ is the pump-controlled parametric coupling and has dimensions of an inverse time. The operators $\hat a$ and $\hat b$ annihilate photons in the signal and idler modes, respectively. In the resonant interaction picture, the free phase rotation is absorbed into these operators. For a constant coupling $\gamma(t)=\gamma$, the evolution equation and its vacuum initial condition are
\begin{equation}
\frac{d\lambda_k}{dt}
=
\gamma(1-\lambda_k^2),
\qquad
\lambda_k(0)=0,
\label{eq:optical-lambda-evolution}
\end{equation}
with solution
\begin{equation}
\lambda_k(t)=\tanh(\gamma t),
\qquad
q_{\mathrm{opt}}(t)=|\lambda_k(t)|^2=\tanh^2(\gamma t).
\label{eq:optical-lambda-solution}
\end{equation}
The reduced density matrix is
\begin{equation}
\rho_a^{\mathrm{opt}}(t)
=
\left[1-q_{\mathrm{opt}}(t)\right]
\sum_{n=0}^{\infty}
q_{\mathrm{opt}}^n(t)|n\rangle\langle n|.
\label{eq:optical-reduced-density-matrix}
\end{equation}
Using the definitions
\begin{equation}
\begin{aligned}
\mu_{\mathrm{opt}}
&=\operatorname{Tr}\!\left[(\rho_a^{\mathrm{opt}})^2\right],
&
S_{L,\mathrm{opt}}
&=1-\mu_{\mathrm{opt}},
\\
S_{2,\mathrm{opt}}
&=-\ln\operatorname{Tr}\!\left[(\rho_a^{\mathrm{opt}})^2\right],
&
S_{\mathrm{vN},\mathrm{opt}}
&=-\operatorname{Tr}\!\left(\rho_a^{\mathrm{opt}}\ln\rho_a^{\mathrm{opt}}\right),
\end{aligned}
\label{eq:optical-diagnostic-definitions}
\end{equation}
the final results are
\begin{equation}
\begin{aligned}
\mu_{\mathrm{opt}}(t)
&=
\frac{1-|\lambda_k(t)|^2}{1+|\lambda_k(t)|^2},
&
S_{L,\mathrm{opt}}(t)
&=
\frac{2|\lambda_k(t)|^2}{1+|\lambda_k(t)|^2},
\\
S_{2,\mathrm{opt}}(t)
&=
\ln\!\left[
\frac{1+|\lambda_k(t)|^2}{1-|\lambda_k(t)|^2}
\right],
&
S_{\mathrm{vN},\mathrm{opt}}(t)
&=
-\ln\!\left[1-|\lambda_k(t)|^2\right]
-\frac{|\lambda_k(t)|^2}{1-|\lambda_k(t)|^2}
\ln|\lambda_k(t)|^2.
\end{aligned}
\label{eq:optical-diagnostic-results}
\end{equation}
Appendix~\ref{app:lambda-framework} gives the intermediate trace and series calculations leading to Eq.~\eqref{eq:optical-diagnostic-results}. Thus, the same $\lambda_k$-based description applies to dynamics generated by a quadratic two-mode squeezing Hamiltonian.

\section{Numerical solutions of the cosmological and optical models}
\label{sec:numerical-solutions}

The numerical analysis applies the $\lambda_k$ framework to paired Gaussian states that admit a Bogoliubov-transformed two-mode representation generated by a Hermitian quadratic bosonic Hamiltonian, with $|\alpha_k|^2-|\beta_k|^2=1$ and $\lambda_k=\beta_k/\alpha_k$. The Bogoliubov representation absorbs the deformation of the paired-state amplitudes into the coefficients $\alpha_k$ and $\beta_k$, thereby converting the wave function into a form whose reduced-state spectrum is controlled directly by $q_k=|\beta_k/\alpha_k|^2=|\lambda_k|^2$. The Riccati equation replaces the coupled evolution of the squeezing amplitude and angle, and the purity, linear entropy, R\'enyi-2 entropy, and von Neumann entropy follow from $q_k$. The complex Riccati equations are integrated with the adaptive DOP853 algorithm \cite{virtanen2020scipy,hairer1993solving}. The standard limiting cases considered below test the numerical implementation, while the parameter-dependent solutions examine its response to a time-dependent cosmological background or optical drive.

\subsection{Cosmological model}
\label{subsec:cosmological-numerical-solutions}

For the cosmological model, we solve
\begin{equation}
\frac{d\lambda_k}{d\eta}
=
-\frac{1}{\eta}\left(1-\lambda_k^2\right)
-2i\sqrt{k^2+\frac{\alpha}{\eta^2}}\,\lambda_k,
\qquad
\lambda_k(-1000)=0,
\label{eq:sec6-cosmological-evolution}
\end{equation}
with $k=1$ and $\alpha=1/4-\nu^2$. The numerical interval is $-1000\leq\eta\leq-1$, and the four cosmological curves correspond to $\nu=0.5,1,1.5,$ and $2$. An additional standard two-mode squeezed vacuum (TMSV) curve is included as an ideal resonant benchmark. In the figures, the final numerical slice $\eta=-1$ is denoted by $0$ on the horizontal axis to display the evolution endpoint in the convention adopted for the plots.

The standard TMSV benchmark is defined by removing the free phase-rotation term while retaining the same cosmological squeezing drive $g_k(\eta)=-1/\eta$. With $\eta_i=-1000$, its squeezing parameter and Bogoliubov ratio are
\begin{equation}
r_{\mathrm{TMSV}}(\eta)
=
\int_{\eta_i}^{\eta}-\frac{d\eta'}{\eta'}
=
\ln\!\left(\frac{|\eta_i|}{|\eta|}\right),
\qquad
\lambda_{\mathrm{TMSV}}(\eta)
=
\tanh r_{\mathrm{TMSV}}(\eta).
\label{eq:sec6-standard-tmsv-benchmark}
\end{equation}
This curve is not an additional value of $\nu$ and does not represent another cosmological background. It is an ideal resonant SU(1,1) reference that shows how much squeezing would accumulate from the same interaction profile in the absence of phase rotation.

\begin{figure}[t]
	\centering
	\includegraphics[width=0.72\linewidth]{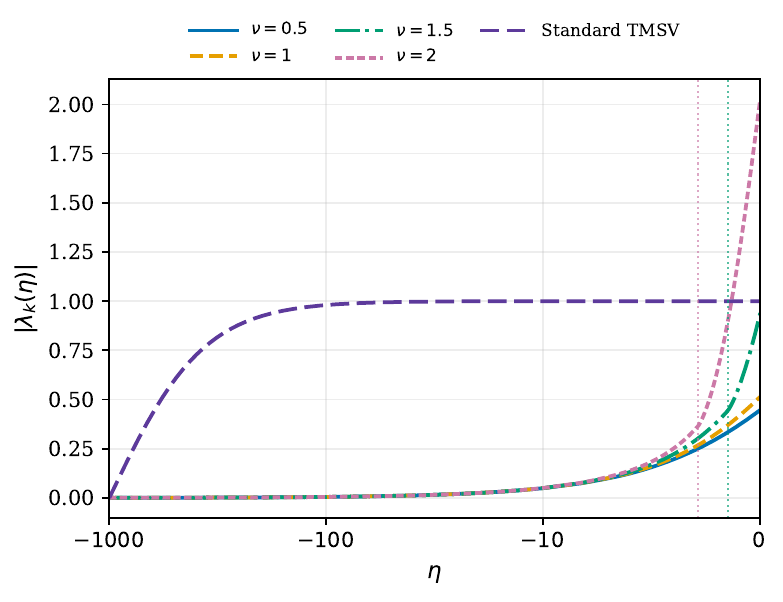}
	\caption{Evolution of $|\lambda_k|$ for the cosmological model and the standard TMSV benchmark. The parameters are $k=1$, $\nu=\{0.5,1,1.5,2\}$, $\alpha=1/4-\nu^2$, $-1000\leq\eta\leq-1$, and $\lambda_k(-1000)=0$; the endpoint $\eta=-1$ is displayed as $0$. The TMSV reference uses $\omega_k=0$, $g_k=-1/\eta$, and $\eta_i=-1000$, whereas $\nu=0.5$ gives the fixed-frequency case $\alpha=0$ and $\omega_k=k$.}
	\label{fig:sec6-cosmo-lambda}
\end{figure}

\begin{figure}[t]
	\centering
	\includegraphics[width=0.72\linewidth]{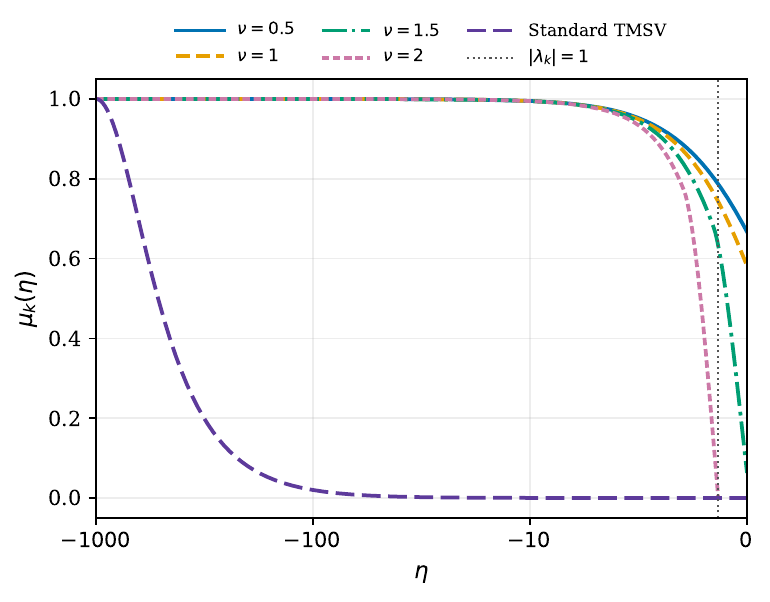}
	\caption{One-mode purity for the cosmological model and the standard TMSV benchmark. The parameters are $k=1$, $\nu=\{0.5,1,1.5,2\}$, $\alpha=1/4-\nu^2$, $-1000\leq\eta\leq-1$, and $\lambda_k(-1000)=0$; the endpoint $\eta=-1$ is displayed as $0$. The TMSV reference uses $\omega_k=0$, $g_k=-1/\eta$, and $\eta_i=-1000$.}
	\label{fig:sec6-cosmo-purity}
\end{figure}

\begin{figure}
	\centering
	\includegraphics[width=0.72\linewidth]{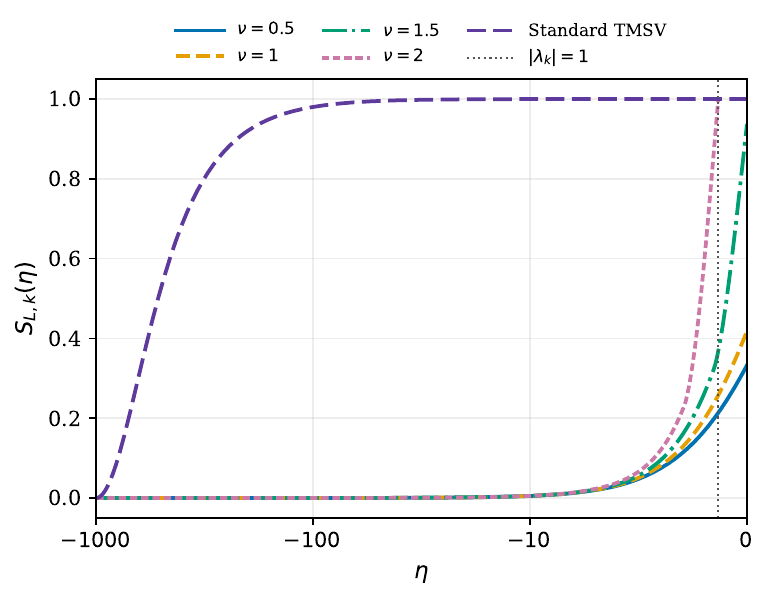}
	\caption{Linear entropy for the cosmological model and the standard TMSV benchmark. The parameters are $k=1$, $\nu=\{0.5,1,1.5,2\}$, $\alpha=1/4-\nu^2$, $-1000\leq\eta\leq-1$, and $\lambda_k(-1000)=0$; the endpoint $\eta=-1$ is displayed as $0$. The TMSV reference uses $\omega_k=0$, $g_k=-1/\eta$, and $\eta_i=-1000$.}
	\label{fig:sec6-cosmo-linear}
\end{figure}

\begin{figure}
	\centering
	\includegraphics[width=0.72\linewidth]{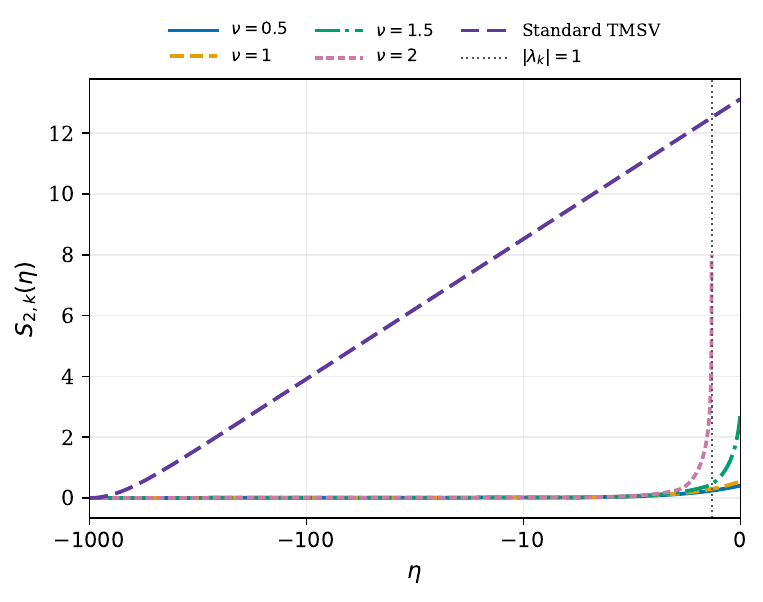}
	\caption{R\'enyi-2 entropy for the cosmological model and the standard TMSV benchmark. The parameters are $k=1$, $\nu=\{0.5,1,1.5,2\}$, $\alpha=1/4-\nu^2$, $-1000\leq\eta\leq-1$, and $\lambda_k(-1000)=0$; the endpoint $\eta=-1$ is displayed as $0$. The TMSV reference uses $\omega_k=0$, $g_k=-1/\eta$, and $\eta_i=-1000$.}
	\label{fig:sec6-cosmo-renyi}
\end{figure}

\begin{figure}
	\centering
	\includegraphics[width=0.72\linewidth]{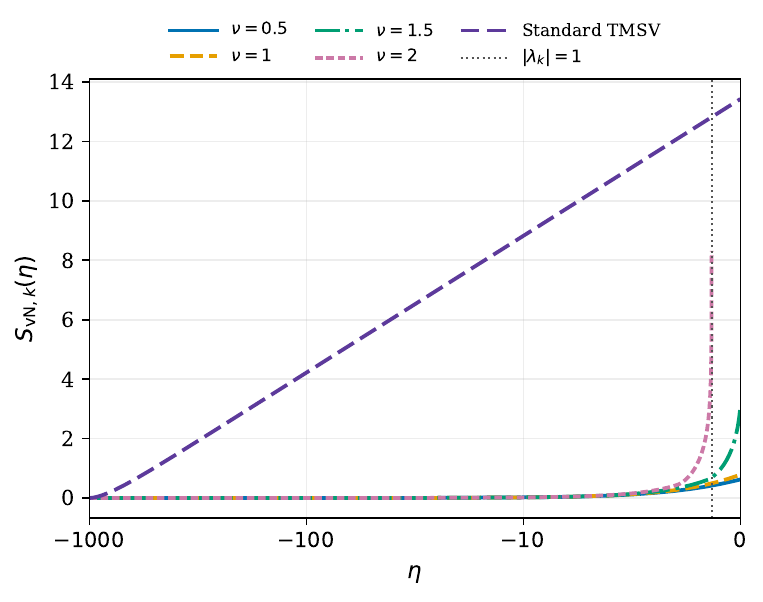}
	\caption{Von Neumann entropy of the reduced $k$ mode for the cosmological model and the standard TMSV benchmark. The parameters are $k=1$, $\nu=\{0.5,1,1.5,2\}$, $\alpha=1/4-\nu^2$, $-1000\leq\eta\leq-1$, and $\lambda_k(-1000)=0$; the endpoint $\eta=-1$ is displayed as $0$. The TMSV reference uses $\omega_k=0$, $g_k=-1/\eta$, and $\eta_i=-1000$.}
	\label{fig:sec6-cosmo-vn}
\end{figure}

The case $\alpha=0$, equivalent to $\nu=0.5$, provides the fixed-frequency benchmark $\omega_k=k$. It retains the cosmological pair-production coupling $g_k=-1/\eta$ but removes the background correction to the free frequency. Unlike the standard TMSV benchmark, the $\nu=0.5$ solution retains the free phase-rotation term $-2ik\lambda_k$; it is therefore the fixed-frequency cosmological reference rather than the ideal resonant TMSV limit. Figure~\ref{fig:sec6-cosmo-lambda} shows that $|\lambda_k|$ is initially negligible and grows as the endpoint is approached. The same growth produces the monotonic decrease in purity in Fig.~\ref{fig:sec6-cosmo-purity} and the corresponding increase in the three entropy measures in Figs.~\ref{fig:sec6-cosmo-linear}, \ref{fig:sec6-cosmo-renyi}, and \ref{fig:sec6-cosmo-vn}. At the final slice, the calculation gives $|\lambda_k|=0.447$, $\mu_k=0.667$, $S_{L,k}=0.333$, $S_{2,k}=0.405$, and $S_{\mathrm{vN},k}=0.625$. This joint behavior is the standard signature of two-mode cosmological squeezing: the complete $(\boldsymbol{k},-\boldsymbol{k})$ state remains pure, while either member becomes mixed after its partner is traced out. The result is consistent with the squeezed-state descriptions of primordial perturbations and their approach to semiclassical behavior in Refs.~\cite{Grishchuk:1990bj,Albrecht:1992kf,Polarski:1995jg}.

For the standard TMSV benchmark, the final slice gives $r_{\mathrm{TMSV}}=\ln 1000=6.908$, $|\lambda_{\mathrm{TMSV}}|=0.999998$, $\mu_{\mathrm{TMSV}}=2.0\times10^{-6}$, $S_{L,\mathrm{TMSV}}=0.999998$, $S_{2,\mathrm{TMSV}}=13.122$, and $S_{\mathrm{vN},\mathrm{TMSV}}=13.429$. The strong separation from the cosmological curves quantifies the suppression of coherent squeezing accumulation by the free phase rotation and the background-dependent mode dynamics.

When $\alpha\neq0$, the term $\alpha/\eta^2$ represents the effective potential generated by the time-dependent cosmological background. In the usual mode equation, $\nu$ encodes the background expansion and the effective mass or slow-roll dependence of the perturbation, so that $k^2-(\nu^2-1/4)/\eta^2$ controls the transition from oscillatory subhorizon evolution to an amplified long-wavelength mode \cite{Mukhanov:1990me,baumann2011physics}. For the values used here, $\alpha$ decreases from $-0.75$ to $-3.75$ as $\nu$ increases from $1$ to $2$. The effective frequency therefore softens earlier, the phase rotation of $\lambda_k$ weakens, and the pair-production term produces a larger amplitude. This explains the ordered curves in all five figures. For example, at the final slice the purity decreases from $0.581$ for $\nu=1$ to $0.063$ for $\nu=1.5$, while the von Neumann entropy increases from $0.786$ to $3.075$. The enhancement of squeezing and reduced-state entropy by a nontrivial background term agrees with established inflationary squeezing analyses and with recent quantum-information studies of cosmological perturbations with modified propagation dynamics \cite{Albrecht:1992kf,Polarski:1995jg,Liu:2026mzz}.
For $\nu=2$, $\omega_k^2$ vanishes at $\eta=-\sqrt{3.75}$, identifying the effective-frequency turning point of this parameter choice.

\subsection{Chirped-pulse optical parametric amplifier}
\label{subsec:optical-numerical-solutions}

\begin{figure}
	\centering
	\includegraphics[width=0.72\linewidth]{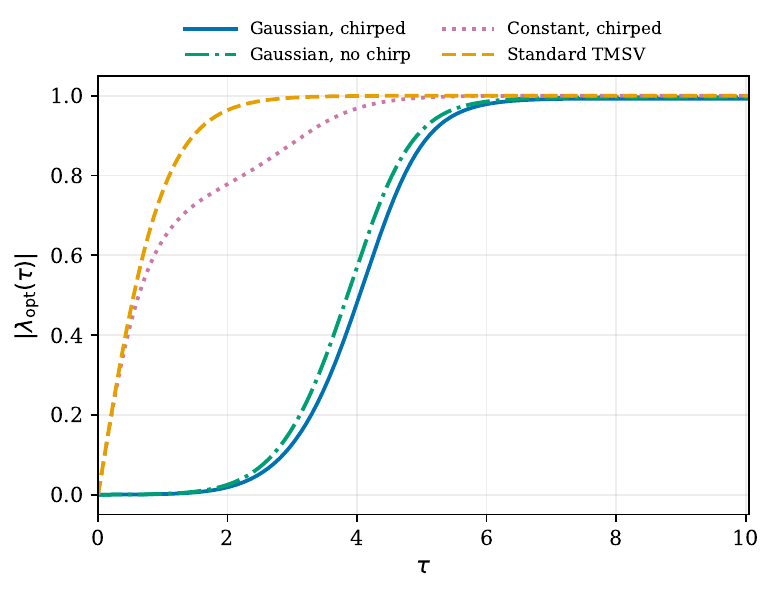}
	\caption{Evolution of $|\lambda_{\mathrm{opt}}|$ for the four optical controls, with $\lambda_{\mathrm{opt}}(0)=0$ and $0\leq\tau\leq10.053$. The Gaussian cases use $s=1.257$ and $\tau_c=5.027$; the chirped cases use $\delta_0=0$ and $\kappa=0.637$. The Gaussian unchirped case has $\kappa=0$, the constant chirped case replaces the Gaussian factor by unity, and the constant resonant TMSV case uses $\kappa=0$ with unit coupling. The dimensional parameters are $\gamma_0/(2\pi)=100\,\mathrm{MHz}$, $\sigma=2\,\mathrm{ns}$, $t_c=8\,\mathrm{ns}$, $t_f=16\,\mathrm{ns}$, $\Delta_0=0$, and $v/(2\pi)=4\times10^{16}\,\mathrm{Hz\,s^{-1}}$.}
	\label{fig:sec6-optical-lambda}
\end{figure}

\begin{figure}
	\centering
	\includegraphics[width=0.72\linewidth]{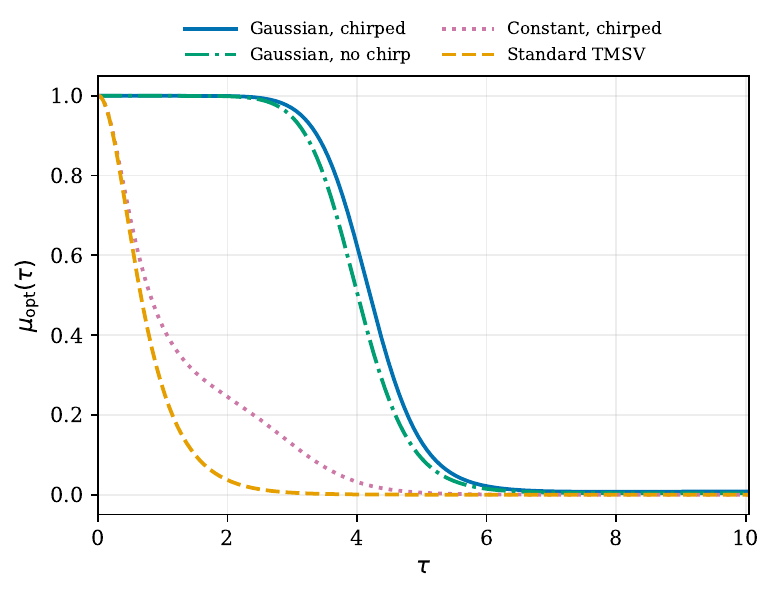}
	\caption{Reduced-mode purity for the four optical controls, with $\lambda_{\mathrm{opt}}(0)=0$ and $0\leq\tau\leq10.053$. The Gaussian cases use $s=1.257$ and $\tau_c=5.027$; the chirped cases use $\delta_0=0$ and $\kappa=0.637$. The Gaussian unchirped case has $\kappa=0$, the constant chirped case uses unit coupling, and the constant resonant TMSV case uses unit coupling with $\kappa=0$. The dimensional parameters are $\gamma_0/(2\pi)=100\,\mathrm{MHz}$, $\sigma=2\,\mathrm{ns}$, $t_c=8\,\mathrm{ns}$, $t_f=16\,\mathrm{ns}$, $\Delta_0=0$, and $v/(2\pi)=4\times10^{16}\,\mathrm{Hz\,s^{-1}}$.}
	\label{fig:sec6-optical-purity}
\end{figure}

\begin{figure}
	\centering
	\includegraphics[width=0.72\linewidth]{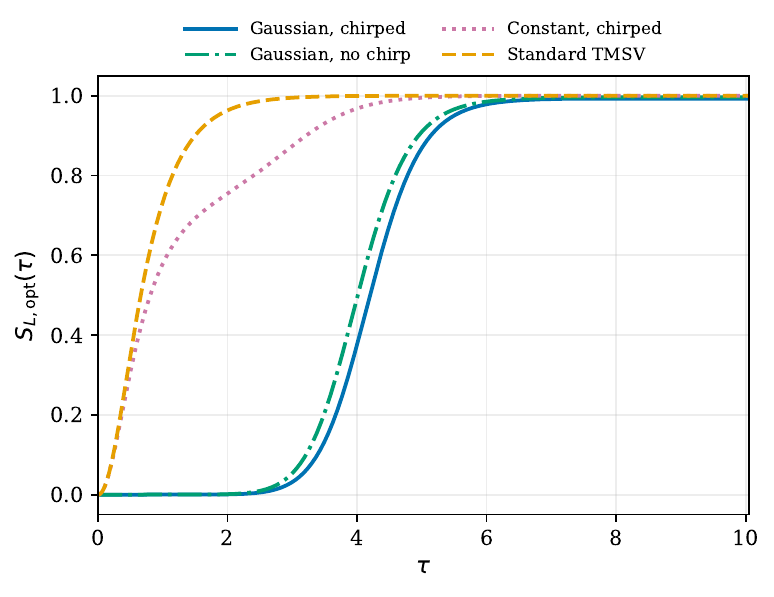}
	\caption{Linear entropy for the four optical controls, with $\lambda_{\mathrm{opt}}(0)=0$ and $0\leq\tau\leq10.053$. The Gaussian cases use $s=1.257$ and $\tau_c=5.027$; the chirped cases use $\delta_0=0$ and $\kappa=0.637$. The Gaussian unchirped case has $\kappa=0$, the constant chirped case uses unit coupling, and the constant resonant TMSV case uses unit coupling with $\kappa=0$. The dimensional parameters are $\gamma_0/(2\pi)=100\,\mathrm{MHz}$, $\sigma=2\,\mathrm{ns}$, $t_c=8\,\mathrm{ns}$, $t_f=16\,\mathrm{ns}$, $\Delta_0=0$, and $v/(2\pi)=4\times10^{16}\,\mathrm{Hz\,s^{-1}}$.}
	\label{fig:sec6-optical-linear}
\end{figure}

\begin{figure}
	\centering
	\includegraphics[width=0.72\linewidth]{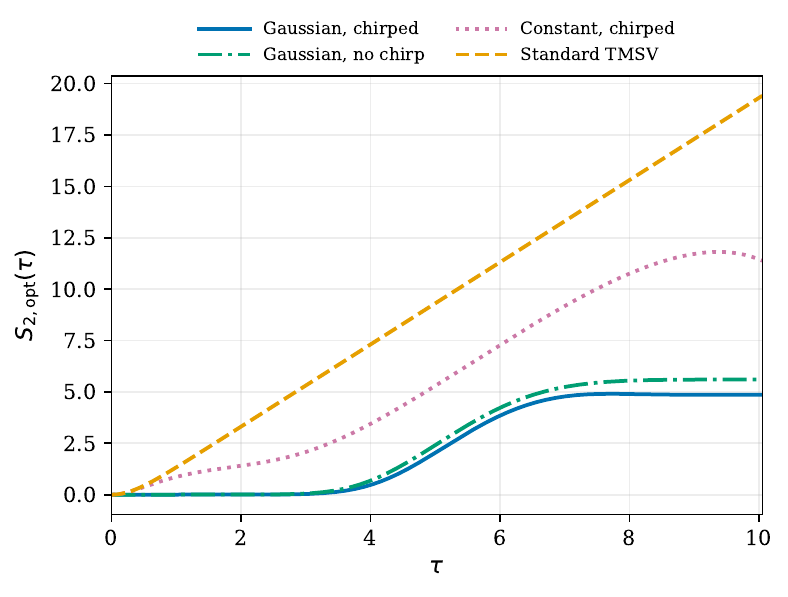}
	\caption{R\'enyi-2 entropy for the four optical controls, with $\lambda_{\mathrm{opt}}(0)=0$ and $0\leq\tau\leq10.053$. The Gaussian cases use $s=1.257$ and $\tau_c=5.027$; the chirped cases use $\delta_0=0$ and $\kappa=0.637$. The Gaussian unchirped case has $\kappa=0$, the constant chirped case uses unit coupling, and the constant resonant TMSV case uses unit coupling with $\kappa=0$. The dimensional parameters are $\gamma_0/(2\pi)=100\,\mathrm{MHz}$, $\sigma=2\,\mathrm{ns}$, $t_c=8\,\mathrm{ns}$, $t_f=16\,\mathrm{ns}$, $\Delta_0=0$, and $v/(2\pi)=4\times10^{16}\,\mathrm{Hz\,s^{-1}}$.}
	\label{fig:sec6-optical-renyi}
\end{figure}

\begin{figure}
	\centering
	\includegraphics[width=0.72\linewidth]{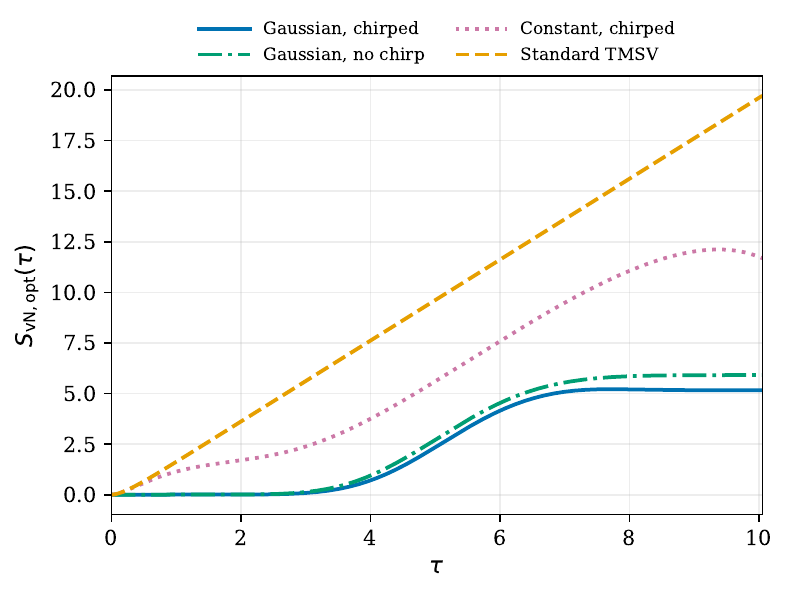}
	\caption{Von Neumann entropy of one optical mode for the four optical controls, with $\lambda_{\mathrm{opt}}(0)=0$ and $0\leq\tau\leq10.053$. The Gaussian cases use $s=1.257$ and $\tau_c=5.027$; the chirped cases use $\delta_0=0$ and $\kappa=0.637$. The Gaussian unchirped case has $\kappa=0$, the constant chirped case uses unit coupling, and the constant resonant TMSV case uses unit coupling with $\kappa=0$. The dimensional parameters are $\gamma_0/(2\pi)=100\,\mathrm{MHz}$, $\sigma=2\,\mathrm{ns}$, $t_c=8\,\mathrm{ns}$, $t_f=16\,\mathrm{ns}$, $\Delta_0=0$, and $v/(2\pi)=4\times10^{16}\,\mathrm{Hz\,s^{-1}}$.}
	\label{fig:sec6-optical-vn}
\end{figure}

The second calculation applies the same procedure to the dimensionless optical equation
\begin{equation}
\frac{d\lambda_{\mathrm{opt}}}{d\tau}
=
\exp\!\left[-\frac{(\tau-\tau_c)^2}{2s^2}\right]
\left(1-\lambda_{\mathrm{opt}}^2\right)
-i\left[\delta_0+\kappa(\tau-\tau_c)\right]\lambda_{\mathrm{opt}},
\qquad
\lambda_{\mathrm{opt}}(0)=0.
\label{eq:sec6-optical-evolution}
\end{equation}
The physical parameters are $\gamma_0/(2\pi)=100\,\mathrm{MHz}$, $\sigma=2\,\mathrm{ns}$, $t_c=8\,\mathrm{ns}$, $t_f=16\,\mathrm{ns}$, $\Delta_0=0$, and $v/(2\pi)=4\times10^{16}\,\mathrm{Hz\,s^{-1}}$. They give $s=1.257$, $\tau_c=5.027$, $\tau_f=10.053$, $\delta_0=0$, and $\kappa=0.637$.Figures~\ref{fig:sec6-optical-lambda}--\ref{fig:sec6-optical-vn} compare four controls: the standard two-mode squeezed vacuum (TMSV) benchmark with constant resonant coupling, a Gaussian envelope without chirp, constant coupling with chirp, and a Gaussian envelope with chirp. The standard TMSV benchmark has $\delta_0=\kappa=0$ and $s\rightarrow\infty$, for which $\lambda_{\mathrm{opt}}(\tau)=\tanh\tau$. Setting only $\kappa=0$ isolates the finite-envelope effect, while replacing the Gaussian factor by unity at fixed $\kappa=0.637$ isolates the effect of the time-dependent detuning under constant coupling.

The orange dashed curve represents the standard TMSV benchmark with constant resonant coupling. It displays the familiar behavior of an undepleted-pump two-mode parametric amplifier: $|\lambda_{\mathrm{opt}}|$ rises immediately toward unity, the reduced-state purity approaches zero, and all entropy measures increase continuously. This agrees with the standard two-photon squeezing formalism and the analytical solution of a resonant parametric interaction \cite{Caves:1985zz,Schumaker:1985zz,walls2008quantum}. The numerical recovery of the $\tanh\tau$ limit verifies the sign, normalization, and information-theoretic conversion used in Eq.~\eqref{eq:sec6-optical-evolution}.

For the physical Gaussian chirped pulse, the coupling is exponentially small far from $\tau_c$, so the state remains close to vacuum during the early part of the evolution. Pair production becomes efficient near the pulse center, where $|\lambda_{\mathrm{opt}}|$ rises sharply and the information measures change over the same time interval. At later times, the Gaussian envelope suppresses the gain while the linearly increasing detuning drives the system away from resonance. At $\tau_f$, the Gaussian chirped case gives $|\lambda_{\mathrm{opt}}|=0.9923$ and $S_2=4.861$, whereas removing the chirp gives $|\lambda_{\mathrm{opt}}|=0.9963$ and $S_2=5.606$. With constant coupling, the chirped case and the standard TMSV benchmark instead give $S_2=11.383$ and $19.413$, respectively. These controls show separately that the finite envelope limits the interaction duration and that the detuning sweep further reduces the accumulated squeezing. The delayed onset and finite plateau are consistent with pulsed parametric amplification, in which the temporal pump profile determines the effective squeezing modes, and with chirped quasi-phase-matched down-conversion, where the detuning sweep controls the amplification bandwidth and accumulated squeezing \cite{Lvovsky:2006vpf,Horoshko:2013zdl,cui2020high}.

The optical comparison complements the cosmological test. The standard TMSV benchmark checks the framework against the exact resonant constant-coupling solution, while the Gaussian chirped pulse introduces two independent physical effects: a localized interaction strength and a time-dependent phase mismatch. Both effects appear in $\lambda_{\mathrm{opt}}$ and propagate consistently to all four reduced-state diagnostics. This common structure reflects the established analogy between cosmological particle creation and laboratory parametric amplification, both of which generate squeezed vacuum states through equivalent pair-production dynamics \cite{Grishchuk:1992tw}. Together with the cosmological results, this agreement shows that, for states admitting the normalized Bogoliubov representation, the proposed calculation separates model-dependent dynamics from the subsequent information-theoretic evaluation: the Hamiltonian fixes the trajectory of $\lambda$, whereas purity and entropy follow from its modulus.

\section{Conclusion and outlook}
\label{sec:conclusion-outlook}
We have recast the paired wave function of a normalized generalized two-mode squeezed state in Bogoliubov form. The evolution considered throughout this work is generated by a Hermitian quadratic Hamiltonian and is not an open-system evolution. Its dynamics is described by the complex ratio \eqref{eq:lambda-bogoliubov-ratio}.  For the general time-dependent quadratic two-mode Hamiltonian \eqref{eq:general-two-mode-hamiltonian}, the Schr\"odinger equation reduces to the Riccati-type evolution equation \eqref{eq:lambda-riccati-general}. 
This single complex variable contains the dynamics conventionally assigned to the squeezing amplitude $r_k$ and phase $\phi_k$. It connects time-dependent pair production directly to reduced-state quantum-information diagnostics, without requiring separate evolution equations for the two squeezing variables. Tracing over the partner mode removes the phase of $\lambda_k$ and gives the geometric one-mode spectrum \eqref{eq:reduced-geometric-spectrum-bogoliubov}. 
The associated diagnostics are shown in Eq. \eqref{eq:lambda-general-diagnostics}.  Model dynamics and information measures are consequently separated as
\begin{equation*}
\begin{aligned}
\bigl\{\omega_k(\eta),g_k(\eta)\bigr\}
&\longrightarrow
\lambda_k(\eta)
\longrightarrow
q_k(\eta)
\\
&\longrightarrow
\bigl\{\mu_k,S_{L,k},S_{2,k},S_{\mathrm{vN},k}\bigr\}.
\end{aligned}
\end{equation*}
Physical inputs change the trajectory of $\lambda_k$, but not the map from $q_k$ to the reduced-state diagnostics. After solving the Riccati equation, one need not reconstruct the reduced density matrix or reevaluate its entropy sums for each model.

The two applications test this separation in different physical settings. In the cosmological model, the fixed-frequency case serves as a reference. An enhanced background effective potential, or a softened effective frequency, changes the free phase rotation and allows pair correlations to accumulate more efficiently. The corresponding increase in $|\lambda_k|$ and $q_k$ as shown in Fig. \ref{fig:sec6-cosmo-lambda}, which it lowers the one-mode purity and raises the three entropies as shown in Figs \ref{fig:sec6-cosmo-purity}, \ref{fig:sec6-cosmo-linear}, \ref{fig:sec6-cosmo-renyi} and \ref{fig:sec6-cosmo-vn}. For the optical model, the resonant constant-coupling limit recovers $\lambda_{\mathrm{opt}}=\tanh\tau$ and checks the Riccati sign, normalization convention, and conversion to the information measures. A finite Gaussian pump concentrates pair production near the pulse center, while the time-dependent detuning limits its later accumulation. The purity and entropies therefore respond after a delay and approach finite plateaus. Despite their different origins, the cosmological background and optical drive enter only through $\omega_k$ and $g_k$, which are all governed by the same $\lambda_k$ equation and reduced-state formulas.

	The Bogoliubov-ratio framework should therefore be understood as a compact dynamical parametrization rather than as a new definition of Gaussian information measures. In the vacuum
	pair sector, the ratio, density-matrix, and covariance-matrix descriptions contain the same information. The advantage of the ratio formulation is that the model-dependent dynamics is
	isolated in a closed Riccati equation and that its modulus and phase transparently separate accumulated occupation from the phase coherence controlling resonant or off-resonant pair
	production. The construction is not restricted to Gaussian states.	Because the Hamiltonian preserves the number difference, the same ratio equation acts in every lowest-weight $SU(1,1)$
	sector. Starting from $|d,0\rangle$ gives an exact non-Gaussian squeezed Fock state whose reduced spectrum is negative-binomial and is completely specified by $(q_k,d)$. The Gaussian formulas are recovered continuously at $d=0$. This result identifies both the extension and its
	boundary: $\lambda_k$ remains a useful group coordinate in more general problems, but an arbitrary non-Gaussian state cannot be reconstructed from this single ratio without additional amplitudes or higher-order correlation data.

\acknowledgments
 S.-C. Liu, L.-H. Liu and B.-C. Li are funded by NSFC grant NO. 12665009 and Hunan Provincial
Department of Education Project NO. 25B0480; H.-Q. Zhang was partially supported by
NSFC with grant NO. 12675057. Y.-B. Zhou is supported in part by the NSFC under Grants No. 12375047, the Hunan Provincial Natural Science Foundation of China under Grants No. 2026JJ30135.

\appendix
\section{Derivations for the $\lambda_k$ framework}
\label{app:lambda-framework}

This appendix gives the derivational details used in Sec.~5. All equations required for the derivation are written explicitly here.

\subsection{Derivation of the evolution equation for $\lambda_k$}

We start from the two-mode state
\begin{equation}
|\psi_k(\eta)\rangle
=
\mathcal{C}_k(\eta)
\sum_{n=0}^{\infty}
\lambda_k^n(\eta)|n_k,n_{-k}\rangle,
\qquad
|\mathcal{C}_k(\eta)|=\sqrt{1-|\lambda_k(\eta)|^2}.
\label{eq:app-lambda-state}
\end{equation}
The phase of $\mathcal{C}_k$ is kept arbitrary, since it represents a global phase and does not affect the reduced density matrix. The quadratic two-mode Hamiltonian is
\begin{equation}
\hat H_k(\eta)
=
\omega_k(\eta)
\left(
\hat a_k^\dagger\hat a_k
+
\hat a_{-k}^\dagger\hat a_{-k}
+1
\right)
+
i g_k(\eta)
\left(
\hat a_k^\dagger\hat a_{-k}^\dagger
-
\hat a_k\hat a_{-k}
\right).
\label{eq:app-general-hamiltonian}
\end{equation}
The state obeys the Schrodinger equation
\begin{equation}
i\frac{\partial}{\partial\eta}|\psi_k(\eta)\rangle
=
\hat H_k(\eta)|\psi_k(\eta)\rangle.
\label{eq:app-schrodinger-equation}
\end{equation}
The left-hand side of Eq.~\eqref{eq:app-schrodinger-equation} acts on the state as
\begin{equation}
i\frac{\partial}{\partial\eta}|\psi_k(\eta)\rangle
=
i\mathcal{C}_k'(\eta)
\sum_{n=0}^{\infty}
\lambda_k^n(\eta)|n_k,n_{-k}\rangle
+
i\mathcal{C}_k(\eta)
\sum_{n=1}^{\infty}
n\lambda_k^{n-1}(\eta)\lambda_k'(\eta)
|n_k,n_{-k}\rangle .
\label{eq:app-schrodinger-left}
\end{equation}
For the right-hand side, the number-conserving part gives
\begin{equation}
\left(
\hat a_k^\dagger\hat a_k
+
\hat a_{-k}^\dagger\hat a_{-k}
+1
\right)
|n_k,n_{-k}\rangle
=
(2n+1)|n_k,n_{-k}\rangle,
\label{eq:app-number-action}
\end{equation}
while the pair creation and annihilation terms give
\begin{equation}
\begin{aligned}
\hat a_k^\dagger\hat a_{-k}^\dagger
|n_k,n_{-k}\rangle
&=(n+1)|(n+1)_k,(n+1)_{-k}\rangle,
\\
\hat a_k\hat a_{-k}
|n_k,n_{-k}\rangle
&=n|(n-1)_k,(n-1)_{-k}\rangle.
\end{aligned}
\label{eq:app-paired-ladder-action}
\end{equation}
Therefore,
\begin{equation}
\begin{aligned}
\hat H_k(\eta)|\psi_k(\eta)\rangle
&=
\mathcal{C}_k(\eta)
\sum_{n=0}^{\infty}
\omega_k(\eta)(2n+1)\lambda_k^n(\eta)
|n_k,n_{-k}\rangle
\\
&\quad
+
i g_k(\eta)\mathcal{C}_k(\eta)
\sum_{n=0}^{\infty}
(n+1)\lambda_k^n(\eta)
|(n+1)_k,(n+1)_{-k}\rangle
\\
&\quad
-
i g_k(\eta)\mathcal{C}_k(\eta)
\sum_{n=1}^{\infty}
n\lambda_k^n(\eta)
|(n-1)_k,(n-1)_{-k}\rangle .
\end{aligned}
\label{eq:app-schrodinger-right-unshifted}
\end{equation}
Re-indexing the last two sums so that all terms multiply the same basis vector $|n_k,n_{-k}\rangle$ yields
\begin{equation}
\begin{aligned}
\hat H_k(\eta)|\psi_k(\eta)\rangle
&=
\mathcal{C}_k(\eta)
\sum_{n=0}^{\infty}
\left[
\omega_k(\eta)(2n+1)\lambda_k^n(\eta)
\right.
\\
&\qquad\left.
+
i g_k(\eta)n\lambda_k^{n-1}(\eta)
-
i g_k(\eta)(n+1)\lambda_k^{n+1}(\eta)
\right]
|n_k,n_{-k}\rangle .
\end{aligned}
\label{eq:app-schrodinger-right}
\end{equation}
Equating Eqs.~\eqref{eq:app-schrodinger-left} and \eqref{eq:app-schrodinger-right} for each paired number state gives
\begin{equation}
i\mathcal{C}_k'\lambda_k^n
+
i\mathcal{C}_k n\lambda_k^{n-1}\lambda_k'
=
\mathcal{C}_k
\left[
\omega_k(2n+1)\lambda_k^n
+
i g_k n\lambda_k^{n-1}
-
i g_k(n+1)\lambda_k^{n+1}
\right].
\label{eq:app-coefficient-matching-direct}
\end{equation}
After division by $\mathcal{C}_k\lambda_k^n$, the relation becomes
\begin{equation}
i\frac{\mathcal{C}_k'}{\mathcal{C}_k}
+
i n\frac{\lambda_k'}{\lambda_k}
=
\omega_k(2n+1)
+
i g_k
\left(
\frac{n}{\lambda_k}
-
(n+1)\lambda_k
\right).
\label{eq:app-matched-coefficients}
\end{equation}
This equation is first read for $\lambda_k\neq0$ and then extended continuously to the vacuum initial condition $\lambda_k=0$. Since it must hold for every non-negative integer $n$, the coefficients of $n$ on both sides are equal:
\begin{equation}
i\frac{\lambda_k'}{\lambda_k}
=
2\omega_k
+
i g_k
\left(
\frac{1}{\lambda_k}
-
\lambda_k
\right).
\label{eq:app-lambda-intermediate}
\end{equation}
Thus the evolution equation for $\lambda_k$ is
\begin{equation}
\lambda_k'
=
g_k(1-\lambda_k^2)-2i\omega_k\lambda_k.
\label{eq:app-lambda-riccati}
\end{equation}
The part of Eq.~\eqref{eq:app-matched-coefficients} that is independent of $n$ fixes only the normalization factor and its global phase. The phase drops out of all one-mode reduced quantities.

\subsection{Reduced density matrix and information-theoretic quantities}

The density operator associated with Eq.~\eqref{eq:app-lambda-state} is
\begin{equation}
\rho_{k,-k}
=
(1-q_k)
\sum_{n,m=0}^{\infty}
\lambda_k^n(\lambda_k^*)^m
|n_k,n_{-k}\rangle
\langle m_k,m_{-k}|,
\qquad
q_k=|\lambda_k|^2.
\label{eq:app-total-density-matrix}
\end{equation}
Tracing over the $-k$ mode and using $\langle m_{-k}|n_{-k}\rangle=\delta_{mn}$ gives
\begin{equation}
\rho_k
=
\operatorname{Tr}_{-k}\rho_{k,-k}
=
(1-q_k)
\sum_{n=0}^{\infty}q_k^n|n_k\rangle\langle n_k|.
\label{eq:app-reduced-density-matrix}
\end{equation}
The eigenvalues of $\rho_k$ are therefore
\begin{equation}
p_n=(1-q_k)q_k^n,
\qquad
\sum_{n=0}^{\infty}p_n=1.
\label{eq:app-reduced-spectrum}
\end{equation}

The purity follows directly from this geometric spectrum:
\begin{equation}
\begin{aligned}
\mu_k
&=\operatorname{Tr}(\rho_k^2)
=\sum_{n=0}^{\infty}p_n^2
=(1-q_k)^2\sum_{n=0}^{\infty}q_k^{2n}
\\
&=\frac{(1-q_k)^2}{1-q_k^2}
=\frac{1-q_k}{1+q_k}.
\end{aligned}
\label{eq:app-purity-derivation}
\end{equation}
The linear entropy and R\'enyi-2 entropy are then
\begin{equation}
S_{L,k}
=1-\mu_k
=\frac{2q_k}{1+q_k},
\qquad
S_{2,k}
=-\ln\mu_k
=\ln\left(\frac{1+q_k}{1-q_k}\right).
\label{eq:app-linear-renyi-results}
\end{equation}
For the von Neumann entropy, substitution of Eq.~\eqref{eq:app-reduced-spectrum} into its definition gives
\begin{equation}
\begin{aligned}
S_{\mathrm{vN},k}
&=-\sum_{n=0}^{\infty}p_n\ln p_n
\\
&=-\ln(1-q_k)
\sum_{n=0}^{\infty}(1-q_k)q_k^n
-\ln q_k
\sum_{n=0}^{\infty}n(1-q_k)q_k^n.
\end{aligned}
\label{eq:app-von-neumann-intermediate}
\end{equation}
Using
\begin{equation}
\sum_{n=0}^{\infty}(1-q_k)q_k^n=1,
\qquad
\sum_{n=0}^{\infty}n(1-q_k)q_k^n
=\frac{q_k}{1-q_k},
\label{eq:app-geometric-identities}
\end{equation}
we obtain
\begin{equation}
S_{\mathrm{vN},k}
=
-\ln(1-q_k)
-\frac{q_k}{1-q_k}\ln q_k.
\label{eq:app-von-neumann-result}
\end{equation}
These expressions show explicitly that the purity and entropies depend only on $q_k=|\lambda_k|^2$. The phase of $\lambda_k$ is removed by the partial trace.

\subsection{Application to the cosmological model}

For the cosmological parametrization used in Sec.~5, the frequency and coupling are
\begin{equation}
g_k(\eta)=-\frac{1}{\eta},
\qquad
\omega_k(\eta)=\sqrt{k^2+\frac{\alpha}{\eta^2}}.
\label{eq:app-cosmological-input}
\end{equation}
Substituting these functions into Eq.~\eqref{eq:app-lambda-riccati} gives
\begin{equation}
\frac{d\lambda_k}{d\eta}
=
-\frac{1}{\eta}(1-\lambda_k^2)
-2i\sqrt{k^2+\frac{\alpha}{\eta^2}}\,\lambda_k.
\label{eq:app-cosmological-lambda}
\end{equation}
The initial condition $\lambda_k(\eta_0)=0$ specifies the initial two-mode vacuum. Once Eq.~\eqref{eq:app-cosmological-lambda} is solved, the reduced spectrum is obtained from
\begin{equation}
q_k^{\mathrm{cos}}(\eta)
=|\lambda_k(\eta)|^2,
\qquad
p_n^{\mathrm{cos}}(\eta)
=
\left[1-q_k^{\mathrm{cos}}(\eta)\right]
\left[q_k^{\mathrm{cos}}(\eta)\right]^n.
\label{eq:app-cosmological-spectrum}
\end{equation}
The one-mode reduced density matrix is therefore
\begin{equation}
\rho_k^{\mathrm{cos}}(\eta)
=
\sum_{n=0}^{\infty}p_n^{\mathrm{cos}}(\eta)
|n_k\rangle\langle n_k|.
\label{eq:app-cosmological-reduced-density}
\end{equation}
The purity is defined by
\begin{equation}
\mu_k^{\mathrm{cos}}(\eta)
=
\operatorname{Tr}\!\left[
\left(\rho_k^{\mathrm{cos}}(\eta)\right)^2
\right].
\label{eq:app-cosmological-purity-definition}
\end{equation}
Using the diagonal spectrum in Eq.~\eqref{eq:app-cosmological-spectrum}, its evaluation reduces to the geometric series
\begin{equation}
\begin{aligned}
\mu_k^{\mathrm{cos}}(\eta)
&=
\sum_{n=0}^{\infty}
\left[p_n^{\mathrm{cos}}(\eta)\right]^2
\\
&=
\left[1-q_k^{\mathrm{cos}}(\eta)\right]^2
\sum_{n=0}^{\infty}
\left[q_k^{\mathrm{cos}}(\eta)\right]^{2n}
=
\frac{1-q_k^{\mathrm{cos}}(\eta)}
{1+q_k^{\mathrm{cos}}(\eta)}.
\end{aligned}
\label{eq:app-cosmological-purity-calculation}
\end{equation}
The linear, R\'enyi-2, and von Neumann entropies are defined, respectively, as
\begin{equation}
\begin{aligned}
S_{L,k}^{\mathrm{cos}}
&=1-\operatorname{Tr}\!\left[
\left(\rho_k^{\mathrm{cos}}\right)^2\right],
\\
S_{2,k}^{\mathrm{cos}}
&=-\ln\operatorname{Tr}\!\left[
\left(\rho_k^{\mathrm{cos}}\right)^2\right],
\\
S_{\mathrm{vN},k}^{\mathrm{cos}}
&=-\operatorname{Tr}\!\left(
\rho_k^{\mathrm{cos}}\ln\rho_k^{\mathrm{cos}}\right).
\end{aligned}
\label{eq:app-cosmological-entropy-definitions}
\end{equation}
For the von Neumann entropy, the key calculation is
\begin{equation}
\begin{aligned}
S_{\mathrm{vN},k}^{\mathrm{cos}}
&=-\sum_{n=0}^{\infty}p_n^{\mathrm{cos}}
\ln p_n^{\mathrm{cos}}
\\
&=-\ln(1-q_k^{\mathrm{cos}})
-\ln q_k^{\mathrm{cos}}
\sum_{n=0}^{\infty}n(1-q_k^{\mathrm{cos}})
(q_k^{\mathrm{cos}})^n.
\end{aligned}
\label{eq:app-cosmological-von-neumann-calculation}
\end{equation}
Here the time argument has been suppressed for compactness. Since
$\sum_{n=0}^{\infty}n(1-q)q^n=q/(1-q)$, the final results are
\begin{equation}
\begin{aligned}
\mu_k^{\mathrm{cos}}(\eta)
&=
\frac{1-|\lambda_k(\eta)|^2}
{1+|\lambda_k(\eta)|^2},
&
S_{L,k}^{\mathrm{cos}}(\eta)
&=
\frac{2|\lambda_k(\eta)|^2}
{1+|\lambda_k(\eta)|^2},
\\
S_{2,k}^{\mathrm{cos}}(\eta)
&=
\ln\left[
\frac{1+|\lambda_k(\eta)|^2}
{1-|\lambda_k(\eta)|^2}
\right],
&
S_{\mathrm{vN},k}^{\mathrm{cos}}(\eta)
&=
-\ln\left[1-|\lambda_k(\eta)|^2\right]
-\frac{|\lambda_k(\eta)|^2}
{1-|\lambda_k(\eta)|^2}
\ln|\lambda_k(\eta)|^2.
\end{aligned}
\label{eq:app-cosmological-diagnostics}
\end{equation}

\subsection{Application to the chirped-pulse optical parametric amplifier}
\label{app:chirped-pulse-opa}

The effective two-mode Hamiltonian in the pump-rotating frame is
\begin{equation}
\hat H_{\mathrm{opt}}(t)
=
\frac{\hbar\Delta(t)}{2}
\left(
\hat a^\dagger\hat a
+\hat b^\dagger\hat b+1
\right)
+i\hbar\gamma(t)
\left(
\hat a^\dagger\hat b^\dagger-\hat a\hat b
\right).
\label{eq:app-optical-interaction-hamiltonian}
\end{equation}
The detuning and coupling are chosen as
\begin{equation}
\Delta(t)=\Delta_0+v(t-t_c),
\qquad
\gamma(t)=\gamma_0
\exp\!\left[-\frac{(t-t_c)^2}{2\sigma^2}\right].
\label{eq:app-optical-input}
\end{equation}
The first term in Eq.~\eqref{eq:app-optical-interaction-hamiltonian} contains the residual detuning after the bare carrier rotation has been removed. Consequently, $\Delta=0$ denotes resonance in the rotating frame and does not imply zero optical carrier frequency.

Identifying $\omega(t)=\Delta(t)/2$ and $g(t)=\gamma(t)$ in Eq.~\eqref{eq:app-lambda-riccati} gives
\begin{equation}
\frac{d\lambda_{\mathrm{opt}}}{dt}
=
\gamma_0
\exp\!\left[-\frac{(t-t_c)^2}{2\sigma^2}\right]
\left(1-\lambda_{\mathrm{opt}}^2\right)
-i\left[\Delta_0+v(t-t_c)\right]\lambda_{\mathrm{opt}},
\qquad
\lambda_{\mathrm{opt}}(t_i)=0.
\label{eq:app-optical-lambda-equation}
\end{equation}
The imaginary term changes the phase of $\lambda_{\mathrm{opt}}$ and, through its competition with pair creation, also changes $|\lambda_{\mathrm{opt}}|$. It therefore cannot be replaced by the bare optical carrier frequency without introducing a physically irrelevant rapid rotation.
Using the dimensionless variables
\begin{equation}
\tau=\gamma_0(t-t_i),
\quad
\tau_c=\gamma_0(t_c-t_i),
\quad
s=\gamma_0\sigma,
\quad
\delta_0=\frac{\Delta_0}{\gamma_0},
\quad
\kappa=\frac{v}{\gamma_0^2},
\label{eq:app-optical-dimensionless-variables}
\end{equation}
and $d/dt=\gamma_0d/d\tau$, Eq.~\eqref{eq:app-optical-lambda-equation} becomes
\begin{equation}
\frac{d\lambda_{\mathrm{opt}}}{d\tau}
=
\exp\!\left[-\frac{(\tau-\tau_c)^2}{2s^2}\right]
\left(1-\lambda_{\mathrm{opt}}^2\right)
-i\left[\delta_0+\kappa(\tau-\tau_c)\right]
\lambda_{\mathrm{opt}},
\qquad
\lambda_{\mathrm{opt}}(0)=0.
\label{eq:app-optical-dimensionless-equation}
\end{equation}
This equation is integrated numerically for finite $s$ or nonzero $\delta_0$ and $\kappa$.
The standard resonant parametric amplifier follows from a controlled limit. Setting $\delta_0=\kappa=0$ removes the effective detuning, while $s\to\infty$ gives
\begin{equation}
\exp\!\left[-\frac{(\tau-\tau_c)^2}{2s^2}\right]
\longrightarrow 1
\label{eq:app-optical-envelope-limit}
\end{equation}
on every finite evolution interval. Equation~\eqref{eq:app-optical-dimensionless-equation} then reduces to
\begin{equation}
\frac{d\lambda_{\mathrm{opt}}}{d\tau}
=1-\lambda_{\mathrm{opt}}^2,
\qquad
\lambda_{\mathrm{opt}}(0)=0.
\label{eq:app-optical-standard-equation}
\end{equation}
Direct integration yields
\begin{equation}
\int_0^{\lambda_{\mathrm{opt}}(\tau)}
\frac{d\lambda'}{1-(\lambda')^2}
=\int_0^\tau d\tau',
\qquad
\operatorname{artanh}\lambda_{\mathrm{opt}}(\tau)=\tau,
\qquad
\lambda_{\mathrm{opt}}(\tau)=\tanh\tau.
\label{eq:app-optical-integration}
\end{equation}
Thus the chirped-pulse model reduces exactly to the standard undepleted-pump two-mode squeezing solution in the resonant, long-pulse limit.
For the general solution of Eq.~\eqref{eq:app-optical-dimensionless-equation}, define
\begin{equation}
q_{\mathrm{opt}}(\tau)=|\lambda_{\mathrm{opt}}(\tau)|^2,
\qquad
p_n^{\mathrm{opt}}(\tau)
=
\left[1-q_{\mathrm{opt}}(\tau)\right]
q_{\mathrm{opt}}^n(\tau).
\label{eq:app-optical-spectrum}
\end{equation}
The reduced state of mode $a$ is
\begin{equation}
\rho_a^{\mathrm{opt}}(\tau)
=
\sum_{n=0}^{\infty}p_n^{\mathrm{opt}}(\tau)
|n_a\rangle\langle n_a|.
\label{eq:app-optical-reduced-density}
\end{equation}
The same geometric-series calculation used in Eq.~\eqref{eq:app-purity-derivation} gives
\begin{equation}
\mu_{\mathrm{opt}}(\tau)
=
\frac{1-q_{\mathrm{opt}}(\tau)}
{1+q_{\mathrm{opt}}(\tau)}
=
\frac{1-|\lambda_{\mathrm{opt}}(\tau)|^2}
{1+|\lambda_{\mathrm{opt}}(\tau)|^2}.
\label{eq:app-optical-purity}
\end{equation}
The entropy measures follow without further dynamical assumptions:
\begin{equation}
\begin{aligned}
S_{L,\mathrm{opt}}(\tau)
&=\frac{2|\lambda_{\mathrm{opt}}(\tau)|^2}
{1+|\lambda_{\mathrm{opt}}(\tau)|^2},
\\
S_{2,\mathrm{opt}}(\tau)
&=\ln\!\left[
\frac{1+|\lambda_{\mathrm{opt}}(\tau)|^2}
{1-|\lambda_{\mathrm{opt}}(\tau)|^2}
\right],
\\
S_{\mathrm{vN},\mathrm{opt}}(\tau)
&=-\ln\!\left[1-|\lambda_{\mathrm{opt}}(\tau)|^2\right]
-\frac{|\lambda_{\mathrm{opt}}(\tau)|^2}
{1-|\lambda_{\mathrm{opt}}(\tau)|^2}
\ln|\lambda_{\mathrm{opt}}(\tau)|^2.
\end{aligned}
\label{eq:app-optical-entropy-results}
\end{equation}

\bibliography{Refs}
\end{document}